\documentclass[preprint]{aastex}

\newcommand{\lya}{Ly-$\alpha$~}
\newcommand{\lyb}{Ly-$\beta$~}

\newcommand{\pcmsq}{cm$^{-2}$}
\newcommand{\ovi}{O~{\sc vi}~}
\newcommand{\civ}{C~{\sc iv}~}
\newcommand{\siiv}{Si~{\sc iv}~}
\newcommand{\siiii}{Si~{\sc iii}~}
\newcommand{\hi}{H~{\sc i}~}
\newcommand{\hii}{H~{\sc ii}~}
\newcommand{\nv}{N~{\sc v}~}
\newcommand{\siii}{Si~{\sc ii}~}
\newcommand{\feii}{Fe~{\sc ii}~}
\newcommand{\cii}{C~{\sc ii}~}
\newcommand{\alii}{Al~{\sc ii}~}
\newcommand{\aliii}{Al~{\sc iii}~}
\newcommand{\oi}{O~{\sc i}~}

\newcommand{\ovii}{O~{\sc vii}~}
\newcommand{\oviii}{O~{\sc viii}~}

\newcommand{\ang}{\AA~}
\newcommand{\mhi}{{\rm H \; \mbox{\tiny I}}}
\newcommand{\movi}{{\rm O \; \mbox{\tiny VI}}}
\newcommand{\mciv}{{\rm C \; \mbox{\tiny IV}}}
\newcommand{\msiiv}{{\rm Si \; \mbox{\tiny IV}}}
\newcommand{\mnv}{{\rm N \; \mbox{\tiny V}}}

\received{2002 March 7}
\begin{document}

\title{Characterizing the Warm-Hot IGM at High Redshift:\\ 
A High Resolution Survey for \ovi at $z=2.5$\altaffilmark{1}}

\author{Robert A. Simcoe\altaffilmark{2}, Wallace L.W. Sargent\altaffilmark{2},
Michael Rauch\altaffilmark{3}}

\altaffiltext{1}{The observations were made at the W.M. Keck Observatory
which is operated as a scientific partnership between the California
Institute of Technology and the University of California; it was made
possible by the generous support of the W.M. Keck Foundation.}
\altaffiltext{2}{Palomar Observatory, California Institute of Technology,
Pasadena, CA 91125, USA; ras@astro.caltech.edu, wws@astro.caltech.edu}
\altaffiltext{3}{Carnegie Observatories, 813 Santa Barbara Street,
Pasadena, CA 91101, USA; mr@ociw.edu}

\begin{abstract}

We have conducted a survey for warm-hot gas, traced by \ovi absorption in
the spectra of 5 high-redshift quasars ($2.2<z<2.8$) observed with 
Keck I/HIRES.
We identify 18 \ovi systems, 12 of which comprise the principal
sample for this work.  Of the remaining six systems, two are
interpreted as ejecta from the QSO central engine, and four have
ionization conditions affected by proximity to the background QSO.
Of the 12 intergalactic \ovi absorbers, 11 are associated 
with complex systems showing strong \lya ($N_{\mhi} > 10^{15.2}$ cm$^{-2}$), 
\civ, and often other lower ionization species.  We do not detect any 
lines that resemble photoionized, enriched gas associated with the 
lowest density regions of the \lya forest ($13.5<\log N_{\mhi}<14.5$).  
Not all of the systems lend themselves to a straightforward
determination of ionization conditions, but in general we
find that they most closely resemble hot, collisionally 
ionized gas found near regions of significant overdensity.  

The extent and gas density of the intergalactic \ovi absorbing regions are
constrained to be $L\le 200$ kpc and $\rho/\bar{\rho}\ge 2.5$.  This
was calculated by comparing the maximum observed \ovi
linewidth with the broadening expected for clouds of different sizes
due to the Hubble flow.  For the median observed value of the Doppler 
parameter $b_{\movi}=16$ km/s, the inferred cloud sizes 
and densities are $L\sim 60$ kpc and $\rho/\bar{\rho}\sim 10-30$.

The clouds have at least two distinct gas phases.  One 
gives rise to absorption in photoionized \civ and \siiv, and has
temperatures in the range $T=20,000-40,000$ K, and overdensities of
$\rho / \bar{\rho}\ge 100$. The second phase is 
traced only in \ovi absorption.  Its temperature is difficult to 
constrain because of uncertainties in the nonthermal contribution to line
broadening.  However, the distribution of upper limits on the
\ovi, \civ, and \siiv temperatures indicates that 
the \ovi thermal structure differs
from that of the other ions, and favors higher temperatures where
collisional ionization would be significant.

The \ovi systems are strongly clustered on velocity scales 
of $\Delta v=100-300$ km/s, and show weaker clustering out 
to $\Delta v = 750$ km/s.
The power law slope of the two-point correlation function is similar to
that seen from local galaxy and cluster surveys, with a comoving
correlation length of $\sim 11h_{65}^{-1}$ Mpc.  
The average Oxygen abundance of the \ovi systems is constrained 
to be $[O/H]\ge -1.5$ at $z\sim 2.5$, about 10 times higher than the 
level observed in the general IGM.  

Two production mechanisms for the hot gas are considered: shock
heating of pre-enriched gas falling onto existing 
structure, and expulsion of material by supernova-driven galactic
winds.  Comparison between the observed numbers of \ovi systems and
expectations from simulations indicates that infall models tend to
overproduce \ovi lines by a factor of $\sim 10$, though this
discrepancy might be resolved in larger, higher-resolution
calculations.   Known galaxy
populations such as the Lyman break objects are capable of producing
the amount of \ovi absorption seen in the survey, provided they drive winds to
distances of $R\sim 50$ kpc.

\end{abstract}

\section{Introduction}\label{sec_intro}

Within the last decade, a picture of the evolving
intergalactic medium has emerged whereby the growth of baryonic
structure is described through the collapse of 
gravitational instabilities 
\citep{cen1994,miraldaescude1996,hernquist1996,zhang1995,petitjean1995}.  
According to this model, baryonic gas 
exists in several different states.  At high redshift, most of the gas is 
found in the \lya forest, which is generally distributed and 
relatively cool at $T\sim 10^4$ K, its temperature governed by
photoionization heating.  Beginning at $z\sim 2.5-3$,
an increasing fraction of the baryons undergo a period of shock heating as
they fall onto large-scale structure.  The cooling timescale for this 
shock-heated phase is long, so by $z=0$ as many as $30\%$ of the
baryons may accumulate in gas with temperatures between 
$10^5-10^7$ K \citep{cen1999,dave2001,fang2001}.  
The remaining $70\%$ of the baryons at the present
epoch have either never been shock heated above $T\sim 10^4$ K, 
or they have cooled much further into highly overdense structures near 
the junctures of filaments.  In these very dense environments 
the effects of local processes begin to play an important role. 

This picture must be incomplete at some level,
since a substantial fraction of the universe is already metal-enriched by
$z=4$ \citep{meyer1987,womble1996,cowie1995,tytler1995},
and the enrichment process is not included in
models relying solely on gravitational instability. Models of metal
absorption lines in a hierarchical scenario \citep{haenhelt1996,
hellsten1997, cen1999, dave2001,fang2001}
have relied either on a very early (``Population III'') 
pre-enrichment phase
or on relatively simple global recipes for calculating 
stellar feedback.
Ongoing metal enrichment at the epoch where we observe the metal
absorption systems may be important, and galactic winds (one of
the possible enrichment mechanisms) have been observed at both low 
and high redshift
\citep{heckman2001b,franx1997,pettini2001}. Moreover there are hints that metal
enriched gas at $z=3$ is turbulent at levels which require energy 
input only $10-100$ Myr prior to the epoch of observation \citep{rauch2001}.

Two of the principal phases of the IGM have been extensively studied 
because they are easily observed in the absorption spectra of high
redshift QSOs as the \lya forest (caused by the cool filaments) \citep{kim1997,
kim2001,rauch1997,mcdonald2001} and 
Lyman limit/Damped \lya systems (caused by the regions of highest overdensity)
\citep[]{prochaska2000}.
However, the hot phase of the IGM with $T>10^5$ K is comparatively
poorly understood, because at such high temperatures the collisional 
ionization of Hydrogen becomes significant, rendering \lya less
effective for tracing structure.  A budget of the content of the 
IGM based on Hydrogen absorption alone will therefore underrepresent the 
contribution of hot gas to the baryon total.

A more accurate account of the hot phase may be made using species with higher
ionization potential than Hydrogen.  The \ovi 1032/1037\ang doublet has long 
been recognized as a prime candidate for this purpose for several reasons
\citep[]{chaffee1986,dave1998,rauch1997}.
First, the intergalactic abundance of Oxygen is higher than that of 
any element other than Hydrogen and Helium.  Second, highly ionized
Oxygen in the form of \ovi, \ovii, or \oviii is among a small number 
of effective tracers for gas in the $T=10^5-10^7$ K range 
typical of shocked environments in cosmological simulations 
\citep[]{cen1999,dave2001,fang2001}.
Among these ionization states, only \ovi is 
visible in ground-based optical spectra of QSOs, at redshifts 
above $z\ge2$.

Further interest in \ovi has revolved around its predicted effectiveness for
tracing heavy elements in the very low density IGM - an environment
very different from the shock-heated one described above.  At 
$T\sim 10,000-40,000$ K, the gas in this diffuse phase is 
too cold for collisional ionization to produce highly ionized species
such as \ovi.  However, its density is sufficiently low (only a few
times the mean) that \ovi may be produced through 
photoionization from the intergalactic UV radiation field.  

Based upon recent simulations, one expects observable
levels of photoionized \ovi to exist in \lya forest lines with 
column densities in the range $13.5 < \log N_{\mhi} < 15.0$ 
\citep[]{hellsten1998,dave1998}.  This \ovi absorption can therefore 
probe the metal content of gas with
densities below the range in which \civ is most sensitive.  Statistical 
studies involving the pixel-by-pixel comparison of optical depths of \lya
and \civ have provided some evidence of widespread enrichment of the IGM 
to even the lowest column densities \citep{songaila1996, ellison2000}, 
and more recently very similar techniques 
have been used to infer the statistical presence of \ovi associated with 
the forest \citep{schaye2000}.  
But to date the number of direct metal line detections 
associatied with $\log N_{\mhi}<14.5$ \lya lines is small, so
the presence of warm photoionized \ovi could help validate the assumption
of widespread enrichment used in some of the simulations described above. 

Considerable attention has attended the recent discovery of
\ovi absorbers in the local neighborhood ($z<0.3$) using 
HST/STIS and the FUSE satellite \citep{tripp2000b,richter2001,sembach2001,savage2002}.
Much of the early interpretation of these results has involved
the difficult job of distinguishing whether particular absorption systems
represent the warm photoionized, or hot collisionally ionized 
variety of \ovi.  Early indications show that the low redshift 
population is mixed,
with a slight majority of collisionally ionized systems.  
Regardless of the physical interpretations of these individual lines, it seems
clear that the baryonic content of the Warm-Hot IGM may be 
significant at low redshift - possibly as much as 30\% of $\Omega_b$.  

In this paper, we describe the results of a survey for \ovi at high
redshift, along the lines of sight to five bright quasars observed 
with the Keck I telescope and HIRES spectrograph.  Our survey covers 
the range $2.2<z<2.8$, which was chosen to optimize the tradeoff 
between signal-to-noise and contamination from the \lya forest.  
We estimate the contribution of warm-hot gas to the total baryon
budget at high redshift, and attempt to identify the physical 
environments of the \ovi systems. 

In Section 2 we describe our observing strategy and methods, and provide
brief descriptions of the properties of individual \ovi systems.
In Section 3 we characterize the physical envronment of \ovi
absorbers, and calculate the contribution of warm-hot gas to the
baryon total.  In Section 4 we discuss possible production
mechanisms for the highly ionized gas in the context of cosmological 
simulations and galaxy feedback on the IGM.  Throughout, we assume a
spatially flat universe with $\Omega_M=0.3$, $\Omega_\Lambda=0.7$,
and $H_0=65$ km/s/Mpc.  

\section{Observations and Data}

\subsection{Survey Strategy}\label{sec_strategy}

Detection of the \ovi 1032/1038\ang doublet from the ground presents a 
particular 
challenge, as the 3000\ang atmospheric cutoff (as well as the decreased 
sensitivity of the HIRES spectrograph blueward of 3200\AA) limits searches 
to $z>2$.  At these redshifts, \ovi is buried in both the 
\lya and \lyb forests, whose densities increase rapidly with 
increasing redshift for $z>1.3$ \citep{kim2001}.  The competition between 
increasing signal to noise ratio toward the red and decreasing contamination 
toward the blue led to our selection of $2.2<z<2.8$ as a target range for the 
survey.  The lower bound was set by instrument sensitivity, and the upper 
bound was derived through consideration of existing studies of line densities 
in the \lya forest \citep[]{kim1997, kim2001, simcoe2002}.  
The 
availability of UV bright QSOs in the spring observing season led us to the
selection of the five sightlines listed in Table 1, with $2.32<z_{em}<2.83$.

Recent observations resulting in the discovery of \ovi in the low
redshift universe \citep{tripp2000b,richter2001,sembach2001,savage2002}
were largely motivated by the desire to trace the hitherto undetected
warm-hot (i.e. $10^5-10^7$ K) component of the IGM, which should be 
most prominent at low
redshift.  However, a major factor in the success
of these surveys is the low level of contamination from the \lya
forest near $z=0$.  In particular, below $z=0.17$ the 
observed wavelength of the \ovi doublet is below 1216\AA, so the
only \hi contamination comes from higher order Lyman
transitions at higher redshift.  These are less numerous than \lya
and can often be easily removed by fitting the corresponding \lya profile.
The surveys cited above generally are limited to $z\le0.3$, so a
portion of their pathlength has no \lya forest absorption, with the
remainder showing contamination at levels $2-3$ times lower than at $z=2.5$.

While the increase in \lya 
forest absorption makes the detection of unblended \ovi more difficult at 
high redshift, the large absorption pathlengths sampled by the more  
distant surveys offset this effect to some extent.  For a given redshift 
interval $dz$, the cosmology-corrected absorption pathlength 
\begin{equation}
\label{eqn_dX}
dX =
{{(1+z)}\over{\sqrt{\Omega_M(1+z)+{{\Omega_\Lambda}\over{(1+z)^2}}}}}~dz, ~~~~(\Omega=1)
\end{equation}       
is $3.3$ times longer at $z=2.5$ than at $z=0$ for 
$(\Omega_M, \Omega_\Lambda)=(0.3,0.7)$.  
High redshift surveys are therefore less clean, but
sample a much larger volume than their low redshift counterparts.  
The total pathlength of our survey is $\Delta z=2.12, \Delta X=6.90$ for the 
above cosmology, which represents a sixteen-fold increase over the total
distance of all published local surveys for \ovi \citep{savage2002}.
In practice we can detect \ovi over $\sim 40\%$ of this range because
of blending with \lya forest lines (See section \ref{sec_xsection}).

\subsection{Observations}\label{sec_observations}

Four of the five objects in our program were observed in excellent 
conditions over the nights of UT 16-17 March 2001, using the Keck I telescope 
and HIRES spectrograph with the UV blazed cross-disperser installed.  
Q1700+6416 had been observed on UT 24 March 1998 and UT 13-15 April 1999
in the blue at high S/N and was therefore also included in our sample.  
All exposures were taken through a 0.86 arcsecond slit providing a measured 
resolution of 6.6 km/s, and the slit was fixed at the parallactic angle 
throughout.

The raw CCD frames were processed and traces extracted to produce 
2-D echelle spectra using the ``makee'' reduction package written 
by Tom Barlow.  Continua were then fit to the individual exposures on an
order-by-order basis, and the unity normalized spectra were
combined with inverse variance weighting onto a common wavelength
scale.  When combining the data, we also included data
taken previously with HIRES using the red cross-disperser.  The
final added spectra have typical S/N ratios between 20-30 per pixel
(35-50 per resolution element) in our \ovi redshift window, as well as
complete coverage of \lya, \lyb, and several other highly ionized species 
such as \nv, \siiii, \siiv, and \civ at even higher signal-to-noise ratios.  

\subsection{Identification of \ovi systems}\label{sec_identification}

The reduced spectra were searched by eye for the 
\ovi doublet at the correct wavelength separation and optical depth ratio,
to create an initial sample of candidate \ovi systems.  
At this stage, we did not subject our search to the constraint
that there be absorption from any other ions at the same redshift as \ovi.  
Each potential system from this list was then fit as a blend of
Voigt profiles using the VPFIT software package
to verify that the profile shapes were adequately matched
in the 1032\ang and 1037\ang components.  Other ions identified at the same
redshift were fit in a similar fashion.  In the case of \hi, we
included as many transitions as possible from the Lyman series in our
fits.  Typically this included at least Ly$\alpha,\beta,\gamma$, but
in come cases reached up to Ly-11.  

This procedure resulted in the identification of 24 pairs of lines whose 
absorption properties are consistent with those of \ovi.  However,
given the density of the \lya forest at our working
redshift, one must carefully consider the possibility of contamination 
due to chance \lya pairs masquerading as \ovi.
Previous searches for \ovi lines in HST/FOS spectra \citep{burlestytler1996} 
found a high rate of chance coincidences in monte carlo simulations, 
but our increase in resolution by a factor of $\sim30$ should 
significantly decrease this source of false positive identifications.  

To estimate the amount of contamination in our initial sample of 25 systems,
we performed a second search of the reduced spectra for pairs of lines 
that are identical to the \ovi doublet in every way except that the optical 
depth ratio of the two components is reversed.  This test should
be robust, as it identically reproduces the \lya forest 
contamination, clustering, metal contamination, and S/N properties of the 
real search process.
Using this method, we were able to identify 7 ``false'' systems that met the
reversed doublet criteria.  Our ``true'' \ovi sample 
contains over three times as many identifications as this false
sample, which 
implies that the true sample is dominated by real \ovi detections
rather than spurious pairings of \lya forest lines.  With no other input to the 
search/selection criteria, we expect the \ovi sample to be contaminated 
by false pairs at the $\sim 25-30\%$ level.  However, it is possible 
to improve significantly on this figure through consideration of 
the properties of the individual systems we have detected.  

In particular, the false systems share the property that they are not 
found near environments populated by other heavy elements,
or even \lya in many cases.  Physically, one expects to find \ovi
absorption only in reasonably close velocity alignment with \lya absorption.  
In the case of low-density photoionized gas, the alignment should be extremely
close: even for metallicities as high as $\frac{1}{10}Z_\sun$ the \ovi column
density can exceed the \hi column by no more than $\sim 60\%$ in
photoionization equilibrium.  
Since the oscillator
strength for \lya is 3.2 times larger than that of \ovi, photoionized \ovi 
should always be accompanied by \lya in exact velocity alignment at similar 
or greater optical depth.  Collisional ionization equilibrium
calculations indicate
that it is possible for very hot gas to produce absorption in \ovi without 
strong \lya.  However, such systems are unlikely to be found in isolation, as 
some heating mechanism is required to produce and sustain the 
high temperatures necessary for \ovi production.  The most likely
sources are supernova-driven galactic outflows \citep{lehnert1996},
shock-heated gas falling onto large scale filaments
\citep{cen1999,dave2001, fang2001}, or hot gas associated with galaxy
groups or proto-clusters \citep{mulchaey1996}.  
Both of these processes are expected to occur near regions of high 
overdensity - the star-forming galaxy in the case of outflows, and filaments
of $\rho/\bar{\rho}=10-100$ in the large scale structure 
scenario.
Furthermore, both the 
outflow and the infall processes are characterized by velocities of 
$\sim100-500$ km/s.  In such systems, one would expect to see a broad \ovi 
component separated by $\Delta v<500$ km/s from a strong \lya system
that also shows 
\civ and possibly other heavy element species.  We therefore add another
criterion to the selection process for \ovi involving proximity to \lya
and \civ absorption for the two likely physical scenarios discussed above.  
In the photoionized case, we expect to find \lya of similar or greater 
strength than \ovi in extremely close velocity
alignment, and in the collisionally ionized case we expect to see \lya and 
\civ within $~500$ km/s of any isolated \ovi absorption.  
We have enforced these 
criteria by considering only those systems which are either 1) paired with 
saturated \lya at $\Delta v_{\mhi-\movi} < 50$ km/s, or 2) located within 1000 
km/s of a system showing \civ absorption.
 
Application of these two criteria to the ``reversed doublet'' pairs resulted 
in the elimination of four of the 7 false systems.  These four 
all showed $\Delta v_{\mciv-\movi} > 1600$ km/s with no nearbly \lya.
Of the remaining three false systems,  one has
$\Delta v_{\mhi-\movi} < 50$ km/s but no corresponding \civ, and two
have $\Delta v_{\mciv-\movi}=200$ and $675$ km/s.
In contrast, for the sample of 24 real potential \ovi systems, 16  
are located at $\Delta v_{\mciv-\movi} < 1000$ km/s, and 14 of these 
show $\Delta v_{\mciv-\movi} < 100$ km/s
{\em and} $\Delta v_{\mhi-\movi} < 100$ km/s.  Six potential \ovi
systems were eliminated because of failure to meet the criteria 
outlined above.  This number
is encouraging, as it closely matches the number of false systems 
(7) detected in the reversed doublet search.  

We conclude that
the selection of \ovi candidate systems based on doublet spacing and ratio, 
subject to constraints on nearby \lya and \civ, is effective at
reducing the amount of contamination from the \lya forest to $\sim 2$ 
objects in
a sample of 20, or $\sim 10\%$.  For the 14 systems with 
$\Delta v_{\mciv-\movi} < 100$ km/s,
the identification as \ovi should be the most secure, while the contamination
may be somewhat worse for the systems showing 
$100 < \Delta v_{\mciv-\movi} < 1000$ km/s.  
Nevertheless, we include these systems in the analysis because of their 
potential physical importance.  Table 2 presents a list of the 18 candidate
systems which survived the selection process, along with a summary of basic
properties derived from the Voigt profile fitting procedure.  The
table is organized by system, with
columns indicating 1) the system redshift, 2) the number of individual \ovi 
components in the system, 3) the total \ovi column density, 4) the number of
corresponding \hi components with $N_{\mhi}>10^{14.0}$, 5) the column
density of the strongest single \hi component, 6) the velocity
separation between the system and the emission redshift of the
background quasar, 7) the velocity separation between the
strongest components of \ovi and \civ, and 8) The figure number
corresponding to the velocity plot for each system.  When measuring 
\ovi column densities, we found that the errors on individual components were
often somewhat large; this was caused by blending between the
components, which opens up a large region of $\chi^2$ space where
adequate fits can be obtained by trading column density between
different lines in the blend.  In these cases the total column density is
much better constrained than would be inferred from the sum of 
the errors for individual components.  It is this better constrained
value that is quoted in Table 2, with accompanying $1\sigma$ errors.
Figure 1 presents stacked velocity plots of each system in several 
ions of interest, overlayed with the best-fit component model.  

\subsection{General Characteristics of Detected \ovi Systems}\label{sec_characteristics}

The properties of our 19 \ovi systems are not entirely
uniform, and while an interpretation of their physical conditions 
is not possible without reference to 
ionization models, there are two classes of systems which can be 
distingiushed
by inspection.  The first group of absorbers in this category contains
systems that are ejected from the background quasar. We find 
two such examples in the data (noted in Table 2), and they are  
characterized by broad absorption that is clearly
matched in the \ovi 1032\ang and 1037\ang profiles, but 
for which the optical depth ratio is too small.  This phenomenon has been 
observed in other high redshift QSOs \citep{barlow1997}, 
and is thought to result from
partial coverage of the continuum source by small, dense clouds close
to the central engine of the QSO.  These clouds
give rise to saturated absorption over the covered
fraction of the central engine, so the doublet ratio approaces unity for
these patches.  In the uncovered region, unattenuated continuum radiation
is able to escape, raising the zero level of the flux to
create the appearance of a pair of unsaturated lines with equal strength.

Another group of absorbers which we distinguish from the general population
contains systems that are found in the immediate vicinity  
of the background quasar ($\Delta v < 5000$ km/s).  The ionization 
environment of these systems should differ significantly from the 
general IGM because of the locally
enhanced UV radiation field \citep{weymann1981}.  Also, since QSOs are likely 
to be found in regions of high overdensity the inclusion
of this class could skew statistical results because of 
clustering effects.  Four systems (also noted in Table 2) fall into this 
class, which we hereafter refer to as the ``proximity'' systems.

After separating out the two populations which are local to the quasar
environment, we are left with a total of 12 true cosmological \ovi  
systems, which we call the ``intergalactic'' sample and 
which will form the principal focus of this paper.
In 10 of these 12 systems, we have detected other highly ionized species 
within $\Delta v < 100$ km/s of \ovi, including at least \civ, but 
usually also  
\siiii, \siiv and other lower ionization species.  These 10 also show 
strong, saturated \lya absorption with $\log N_{\mhi} > 14.55$, 
(for 9 of the 10, $\log N_{\mhi} > 15.2$).  The identification of these 
systems as \ovi appears to be very secure.

For the remaining 2 intergalactic systems, the \ovi absorption 
is significantly 
offset from the nearest large concentration of \civ (by 
$336$ km/s and $772$ km/s).  These systems are much more difficult to 
distinguish from chance \lya associations, as there is no closely aligned
absorption from any ion other than \ovi - including \lya and \nv.  Based on 
our detection of two reversed doublet systems for which $\Delta v < 700$ km/s,
these pairs might seem to be likely candidates for contamination.  However,
the two offset systems differ from the false pairs in that their 
nearest respective \lya systems are exceptionally rich.
One shows a cluster of 5 lines with $\log N_{\mhi}>15.5$ within a 
700 km/s range, and other has an \hi column 
density of $\log N_{\mhi}=16.0$.  Both also exhibit 
extensive metal line absorption including \civ, \siiv, and \siiii.  
No reversed doublet systems were observed in the vicinity of such 
strong \lya and \civ, and 
even the two reversed doublet systems showing 
$\Delta v_{\mciv-\movi}<1000$ km/s did not
have associated \siiv or \siiii.  The proximity
of the offset \ovi candidates to several of the strongest absorption 
line systems in our survey, along with the fact that the offset velocities 
are exactly those expected in the infall/outflow scenarios discussed above,
have led us to consider these systems as highly probable identifications
despite the increased possibility of contamination.  

From inspection of the final sample of 12 intergalactic systems in
Table 2 and Figures 1-18, it is seen that our survey selects primarily strong
absorbers, with saturated \lya 
and associated heavy element ions.  All of the intergalactic systems
show $\log (N_{\mhi})\ge 14.5$ and observable levels of \civ - i.e. 
we have not detected any ``\ovi only'' absorbers in the forest 
(although we do see two
such cases in the proximity sample).  In fact, we have not 
found a single example of an \ovi doublet aligned with a \lya forest 
line in the column density range $13.5 < \log N_{\mhi} < 14.5$.  For
many systems, such weak \ovi lines would not have been detectable
due to variations in the data quality and degree of 
\lya forest blending.  Since we have not attempted to deblend lines
from \lya and higher order \hi until after they have already been
identified, it is possible that we may
be biased against the weakest \ovi lines, since these are the first
lost to blending from the forest.  We have estimated that for lines
with similar $N_{\movi}$ and $b_{\movi}$ to the ones picked out by our
selection criteria, our survey is $\approx 40\%$ complete 
(see Section \ref{sec_xsection} for a detailed discussion of the 
completeness estimation).  
It is likely that the completeness is lower for the weakest systems
- a tradeoff we have made in order to minimize the number of false positive
detections in our sample.

Nevertheless, there are several systems in our
sightlines where the data is of sufficient quality and the spectrum
sufficiently clean of \lya lines that we would expect to detect \ovi,
even at levels nearly an order of magnitude below [O/H]=-2.5 if a 
Haardt \& Madau (1996) shaped
UV background is assumed.  The detailed question of whether the number
of such systems is significant, or whether it is consistent with the
expected patchiness in the cosmic metallicity and/or ionizing radiation field
is more complex and will be addressed in a companion paper.

Subject to the above caveats, the data and selection techniques presented here 
do not confirm the presence of photoionized metal lines in the 
low-density IGM at the [O/H]=-2.5 level  
\footnote[1]{Carswell, Schaye \& Kim (2002) have very recently reported
  the detection of a population of low density, photoionized \ovi
  absorbers near $z\sim 2$ using different selection criteria from
  those presented here.  See the Appendix of this paper for a
  brief comparison of the two methods and results.}.
This result comes as a 
surprise, as previous statistical studies at higher redshift implied a 
widespread distribution of \ovi in association with the forest 
\citep{schaye2000}, and theoretical considerations had also pointed to 
\ovi searches as the most effective way to test the widespread
enrichment hypothesis \citep{chaffee1986,rauch1997,hellsten1998}.  
Our results agree more closely with those of \citet{dave1998}, who
find a significant downward gradient in metallicity toward low density
regions of the IGM at $z\sim 3.25$.  For the remainder of this paper, 
we limit the discussion to the strong systems which have 
been firmly identified in the survey.

\subsection{Observed Properties of Individual Systems}\label{sec_individual_systems}

In this section we present brief summaries of the notable spectral features
in each \ovi system.  We limit the present discussion to actual 
observed properties; further discussion of the absorbers' physical
conditions is given in Section 3.  The systems are organized
by sightline, in order of increasing redshift.

\subsubsection{Q1009+2956: $z=2.253$ (Figure 1)}\label{sec_1009_2.25}
Our first \ovi candidate is located in the vicinity of a strong Lyman limit
system (LLS, $\log N_{\rm \mhi}=17.8$).  Both \lya and \lyb were used 
to fit for
the \hi column density; higher order Lyman transitions could not be
used due to the presence of another Lyman Limit system at higher
redshift.  Complex chemical absorption containing \civ , \siiv , and
\siiii and spanning roughly 200 km/s is associated
with the strongest \hi absorption.  The velocity
structure in these ions is closely aligned, although the \civ/\siiv ratio 
varies across the profile, suggesting a spatial variation in ionization
conditions.  Our Voigt profile fits indicate that the alignment 
and relative widths of \civ and \siiv are consistent with pure
thermal broadening of the profile at $T_{C,Si} = 0.8-1.2\times 10^5$ K 
(reflecting the range for different individual components).  

The \ovi profile differs markedly from those of \siiv and \civ, and
is characterized by a broad trough with little
substructure.  This smooth, blended nature causes the Voigt fit parameters 
to be poorly constrained - particularly the line widths.  For this reason,
and also because of blended \lya forest absorption
in the \ovi 1037 \ang profile (seen to the blue in Figure 1a) any
detailed physical conclusions about the \ovi gas remain tentative.  
However, it is still evident from inspection of the profiles that the 
small amount of structure in the \ovi line does not mirror that of
\civ and \siiv.  
Taking the best-fit Voigt
parameters at face value, the measured limits on the \ovi temperature 
range from $T_{\movi}\le 0.2-2.1\times 10^6$ K for various components,
or roughly an order of magnitude above the measured temperatures for
\civ and \siiv.  

\subsubsection{Q1009+2956: $z=2.429$ (Figure 2)}\label{sec_1009_2.42}

This system is dominated by a single strong \hi component, whose
column density was measured at $\log(N_\mhi)=17.687$ using
\lya,$\beta,\gamma$ and $\delta$.  As expected for such a strong \hi
system, significant absorption is present in lower ionization species
including \cii, \siii, and \alii.  The \civ, \siiv, and \siiii profiles are
closely aligned, and contain at least one very narrow ($b_{\mciv}=4.8$
km/s) component that aligns with \cii and implies a temperature of
$T<14,000$ K.  Similar linewidths for the \siiv and \civ profiles
indicate that their broadening may be largely non-thermal, and the
temperatures even lower.

The \ovi profile contains two subcomponents, neither of which align
in velocity with the lower ionization lines.  Their line widths of
$b_{\movi}=22.2,24.7$ km/s imply upper limits of $T_{\movi}<5\times
10^5$ K.  

\subsubsection{Q1009+2956: $z=2.606$ (Figure 3)}\label{sec_1009_2.60}
This system consists of a single, isolated \lya line  
with associated \ovi, but no absorption from any other heavy elements.  
The \hi resembles a typical
\lya forest line, at $\log(N_{\mhi})=14.442$ and $ b_{\movi}=25.6$ km/s.
A single \ovi line is detected at the $\ge 3\sigma$ level, and is
measured to have $\log(N_{\movi})=12.709$ and $b_{\movi}=8.0$
km/s ($T_{\movi}\le 6.2\times 10^4$ K).  This is an example of the type
of system we had expected to detect in large numbers in our survey.
However, this particular absorber is located only 3050 km/s from the
background quasar and its ionization state is likely to be 
affected by UV radiation from the QSO.  It has therefore been
grouped with the ``proximity'' sample and not included in our
discussion of cosmological \ovi absorbers.  However, its detection illustrates 
that the survey is sufficiently sensitive to 
uncover \ovi only systems, even
though we have found none in the more tenuous regions of the IGM.

\subsubsection{Q1442+2931: $z=2.438$ (Figure 4)}\label{sec_1442_dla}
This \ovi absorption is associated with a weak damped \lya (DLA) system.
The damping wings of the profile spread over several echelle orders,
complicating attempts to determine the exact \hi column density.
However, based on the presence of the modest damping wings, and a  
relatively weak saturation level in the core by DLA standards, we
estimate that $\log N_{\mhi}\sim 19.0-20.0$.  A rich metal line
structure is detected both in low ionization (\oi, \siii, \feii,
\alii, \cii) and high ionization (\civ, \siiv) species.
%
%
The kinematic spread of the lowest ionization gas (e.g. \feii, \oi) 
spans a range of $\sim 200$ km/s and is centered near the strongest
\hi absorption.  The highest ionization gas (\civ, \siiv, \ovi) 
also spans a $\sim 200-250$ km/s velocity interval, but is offset 
$\sim 200$ km/s to the red of the low ionization species.  Several intermediate
ionization lines (\siii, \cii, \aliii) bridge the velocity range
between the high and low ionization species, sharing absorption
components with both varieties of gas.  

The \ovi kinematics strongly resemble those of \civ, where a strong absorption
trough at $z=2.390$ is flanked by two weaker structures at $\pm 100$
km/s.  A detailed comparison of the central regions of the profiles is
not possible because of saturation in the \civ core.  

\subsubsection{Q1442+2931: $z=2.623$ (Figure 5)}\label{sec_1442_2.62}

This \ovi absorption is associated with a pair of strong \hi systems,
seen at $\sim \pm 300$ km/s in Figure 1f.  The \hi at $-300$ km/s
splits into three strong components in higher order Lyman lines, and
we measure these to have column densities of $\log(N_{\mhi})=15.3,15.3,$ and
$14.6$.  The group at $+300$ km/s is dominated by a single line with  
$\log(N_{\mhi})=15.8$.  The whole complex is located near the
background quasar at $\Delta v \approx 3100$ km/s, 
so we have grouped it in the proximity sample and excluded it from our
cosmological statistics.  

This system was originally identified as an \ovi absorber based on the
narrow line located near +250 km/s in the figure.  Though this 
line is blended with \lya absorption from the forest in the 1032 \ang 
component, a sharp core is quite visible in the profile that mirrors 
the shape of the uncontaminated \ovi 1037 A line.  The width of this 
core is relatively small at $b=7.41$ km/s,
and implies an upper limit of $T_{\movi}\le 5.27\times 10^4$ K.  
A second, much broader component ($b=19$ km/s) is also seen in the red wing
of this line.  \civ is detected in the same velocity range as the sharp
\ovi component, further securing the \ovi identification for the system. 
The \ovi and \civ may come from the same gas, as the
redshifts are identical to within the $1\sigma$ errors, and the
velocity widths are also within $1\sigma$ of a simulateneous solution
of $T_{\mciv,\movi}=8.5\times 10^4$ K.  The column density ratio for these
lines is $\log(N_{\movi}/N_{\mciv})=0.454$.  

The absorption complex at $-300$ km/s shows a much richer chemical
structure, containing kinematically complex \civ, and also weak \siiv
and \siiii.  We did not originally identify \ovi associated with this
\hi using our search strategy, because we could not be certain
that much of the absorption in the 1037 \ang profile over this
range was not \lya contamination.  However, in light of the nearby sharp
\ovi line discussed in the preceeding paragraph, the strong \hi, \civ,
\siiv, and \siiii, and the good match of the 1032 \ang and 1037 \ang
profiles, we treat this as a possible, or tentative \ovi identification.
If this absorption actually represents \ovi, its
properties are different from those of the \civ and \siiv lines.
Solving simultaneously for the temperature using the \civ and \siiv
linewidths , we find $T_{C,Si}=1.2-2.6\times 10^4$ K, while attempts
to fit the \ovi profile yield linewidths in the range
$b_{\movi}=10.6-32.8$ km/s, or $T_{\movi}\le 1\times 10^5-1\times
10^6$ K.  This system appears not to be unique in its association with 
a group of strong \hi and metal lines rather than a single strong
system (See for example Section \ref{sec_1700_2.43}).

\subsubsection{Q1549+1919: $z=2.320$ (Figure 6)}\label{sec_1549_2.32}

This example contains a broad complex
of \ovi distributed over $\Delta v \sim 500$ km/s.  
The associated \hi is once again not dominated by a single line, but
rather by a 
cluster of three moderately strong lines ($\log N_{\mhi}=15.195,
14.627,14.111$) within a small velocity interval.  
The \ovi in this system was originally identified by the 
presence of two narrow lines seen at the blue end of the profile.
We detect a single weak \civ line associated with this narrow \ovi
component.  Although
the \civ and \ovi do not align exactly in velocity, their doppler
widths both indicate upper limits on the temperature which are quite
cool, in the range $T\le 3.2-7.4\times 10^4$ K.  One of the \ovi lines is
extremely narrow at $1.46$ km/s ($T_{\movi}\le 2000$ K), although this
may be a noise artifact.

A second pair of \civ lines is detected near the strongest \lya line in
the system.  This additional \civ is unusually broad and
featureless - the velocity width of one of the two \civ components is
measured at $b=19.67$ km/s, or $T_{\mciv}\le 2.8\times 10^5$ K.  Such an
environment should be conducive to the production of other very highly
ionized species, including \ovi.  However, we did
not at first identify any \ovi associated with the broad \civ because 
the \ovi 1032 \ang line was blended with a higher redshift \lyb line.  
The contamination was removed by fitting the corresponding \lya profile
to infer the \lyb line strength, revealing the profile shown in the figure.

The shape of the deblended profile suggests that \ovi exists over 
the entire range of \hi absorption.  Most of this \ovi is lacking in
detailed substructure.  Our Voigt profile fits help to characterize
several of the apparently broad features, although the errors
on the fit parameters are significant because of the smoothness of the
profile.  The strongest distinct feature is a broad line near the 
strongest \hi component and wide \civ ($+50$ km/s in Figure 6).  Its 
measured $b_{\movi}=26.43$ implies an upper limit on the temperature of 
$T\le 6.7 \times 10^5$ K.  This matches the high temperature inferred 
from \civ to within $1\sigma$ errors, although the different shapes of
the \ovi and \civ profiles indicate that the absorption 
probably does not arise in the same gas.  

\subsubsection{Q1549+1919: $z=2.376$ (Figure 7)}\label{sec_1549_2.37}

This system is
offset by more than $700$ km/s from the nearest \civ line.
No absorption is seen from any ion including \hi at the \ovi redshift,
so the identification rests on the similarity of the doublet profiles alone.
These are identical to within the noise over most of their length, though 
some blending is present in the 1037\ang component.  

The nearest \hi to this system is an unusual complex, containing
five lines of $\log(N_{\mhi})>14.0$ within a span of $600$ km/s.  No
single line dominates the complex; the strongest (located at $+750$
km/s from the \ovi) is measured at $\log(N_{\mhi})=15.6$.  \civ, 
\siiii, and \siiv are clearly detected in association with the strongest
\hi component, but the \ovi profile suffers from heavy \lya forest blending
in this region.  Using profile fits of the C and Si
ions to solve simulatneously for the gas temperature 
and non-thermal motions, we estimate $T_{C-Si}=2.2\times 10^4$ K.  

Our best fit Voigt profile for the \ovi (based primarily on the unblended 
1032 \ang component) requires two lines, with widths of 
$b=25.24,26.70$ km/s.  The implied upper limits on the \ovi temperature are 
$T_{\movi}\le 6.5\times 10^5$ K - over an order of magnitude higher than 
that of C or Si, although the \ovi measurement is only an upper limit.  
High quality data in the \civ and \nv regions 
($\frac{S}{N}(\mciv)\approx 150, \frac{S}{N}(\mnv)\approx 120$ per pixel) allow
to set extremely strong limits on the non-detection of other ions
associated with \ovi: in particular $\log(N_{\movi}/N_{\mnv})>2.398$ and 
$\log(N_{\movi}/N_{\mciv})>2.717$.  

\subsubsection{Q1549+1919: $z=2.560$ (Figure 8)}\label{sec_1549_2.56}

This system contains a single \ovi line with no associated
\hi absorption, although a strong \hi absorber of $\log(N_{\mhi})=15.219$
is located $336$ km/s to the blue.  The \ovi 1032 \ang line
is blended with interloping \hi, but a distinct narrow component is
still visible in this blend, matching in redshift and doppler parameter
with the clean \ovi 1037 \ang profile.  The measured $b_{\movi}=14.6$ km/s 
translates to an upper limit of $T_{\movi}\le2.0 \times 10^5$ K.  

As in the preceding example, no absorption is seen from any other heavy
elements at the redshift of the confirmed \ovi: we measure limits of
$\log(N(\movi)/N(\mciv))\ge 2.701$ and \\
$\log(N(\movi)/N(\mnv))\ge 2.038$.
However, a significant amount of \civ is associated with the \lya
complex at $-336$ km/s.  In this region,
the detailed structure of the \ovi 1032 \ang profile seems to mirror that
of the \civ, suggesting that there may be further \ovi locked up in
this system.  Indeed, the observed \ovi substructure is almost
certainly not the product of chance \lya forest absorption, as it varies
on velocity scales of $\sim 10$ km/s, whereas characteristic
velocity widths in the \lya forest are $b>20$ km/s.  
Nevertheless, we do not claim a direct detection 
of \ovi in this part of the profile, as the \ovi 1037\ang line is 
saturated by a blended \lya forest line, and further examination of
the 1032\ang component indicates that it suffers from blending as well.
 
\subsubsection{Q1549+1919: $z=2.636$ (Figure 9)}\label{sec_1549_2.63}

This system has a particularly simple chemical and velocity
structure.  The \lya profile is dominated by a single \hi line with 
$\log(N_{\mhi})=15.220$, which places it among the weakest systems in the
survey.  Several other systems contain groups of lines with similar
\hi column densities, but this system is unusual in its
isolation from other strong \hi.

Heavy element absorption is only seen in \civ and \ovi, both of which
are adequately fit by a pair Voigt profile components.  The velocity alignment
between the ions is relatively close, but for both components the \ovi
linewidth exceeds that of \civ despite the fact that \ovi is a heavier
ion than \civ.  The \civ linewidth constrains its temperature to be below
$T_{\mciv}<5.0\times 10^4$ K, while the \ovi linewidth permits
temperatures in the range $T_{\movi} < 1.3,2.6\times 10^5$ K for the
two components.  

\subsubsection{Q1549+1919: $z=2.711$ (Figure 10)}\label{sec_1549_2.71}

This absorber is an example of highly ionized ejecta from
the background QSO, moving at $v_{ej}\approx 10,000$ km/s.  A wide,
sloping absorption profile is seen in \ovi, \civ, and \nv 
(only \nv 1238 \ang is shown, as the 1242 \ang line is contaminated).  
The \hi seen at the same redshift is not likely to be associated with
the ejecta, as it shows no sign of the unusual kinematics that characterize
the high ionization lines.  The principal evidence for the ejection 
hypothesis comes from the doublet ratios of \ovi, \civ, and \nv, 
which are unity over their whole profiles rather than the value
deduced from the oscillator strengths of different transitions (see
Section 2.4 for a more thorough discussion of this phenomenon).
Accordingly, we have excluded this system from our further statistical 
analysis.

\subsubsection{Q1626+6433: $z=2.245$ (Figure 11)}\label{sec_1626_2.24}

This \ovi system - our lowest redshift example - is somewhat difficult 
to interpret due blending and saturation in the \ovi 1032 \ang component,
and a relatively low signal-to-noise ratio ($\sim 7$).  It is associated
with a multicomponent \hi complex of total column density $\log(N_{\mhi})=
15.544$ seen as a single line in \lya, although a fit to \lyb reveals 
several subcomponents.  The strongest of these is 
saturated in both \lya and \lyb, but higher order transitions could not 
be used in the fit because of their poor signal-to-noise ratios.  

Heavy element absorption is seen in \cii, \civ, \siiv, and \siiii. 
In each case, the metals are in two clusters - one associated with the
densest \hi, and a second, weaker group near a smaller \hi line at $-200$
km/s.  The \ovi absorption is only found near the
stronger line, though a weak signal at $-200$ km/s could be masked by noise.
Since the strong \civ profile is saturated in the 1548\ang component, 
we have used only the unsaturated 1551\ang component to measure the
velocity profile.  The individual elements of the \civ, \siiii, and 
\siiv absorption align well in velocity, and the fits indicate a
temperature range of $T=2-8\times 10^4$ K for the Carbon and Silicon gas.  

Because of lower data quality, it is difficult to judge whether the
velocity structure of \ovi matches that of the \civ and \siiv
profiles.  The overall velocity extent appears to be similar, but any
substructure is lost in the noise.  We have therefore
compared two different approaches to measure the system's
properties.  The first of these assumes that the \ovi absorption comes 
from the same gas as \civ, \siiii, and \siiv, and is motivated by the 
similar extent of \ovi and \civ in velocity space.  To test the 
plausibility of this hypothesis, we have
fit a model \ovi profile that contains 8 components with the same
redshifts and $b$ parameters (reweighted by $\sqrt{m_{\mciv}/m_{\movi}}$) 
as the \civ profile.  The column densities were then allowed 
to vary to achieve the best fit.  The \ovi line
strengths measured in this way yield typical ratios of
$\log(N_{\movi}/N_{\mciv})\sim 0.3$ and $\log(N_{\movi}/N_{\msiiv})=1.5-2.2$
for individual absorption components.  We then compared these ratios with
the predictions of ionization simulations
(described in detail in section 3.1.3) to see if they are consistent
with a photoioniation interpretation.  From this exercise we found
that if the \civ and \siiv are produced in the same gas, then the
observed \ovi line strength is much stronger then expected.  We therefore
consider it likely that the \ovi gas is physially distinct from the
\civ and \siiv gas, which is similar to what is seen in most other systems.

Our second approach assumes that the C and Si absorption
are associated, but that the \ovi is contained in a separate phase.
A Voigt profile fit to the \ovi 1037\ang profile requires only
two components to adequately represent the data ($\chi^2_{\nu}=1.02$).  
The fit does not match the 1032 \ang profile exactly, but the discrepancy
can be attributed to forest contamination.  Most of the \ovi absorption is
contained in a single component with $\log(N_{\movi})=14.8$ and
$b_{\movi}=38$ km/s ($T_{\movi}\le 1.4\times 10^6$ K).  This total
column density is actually quite similar to the total column measured 
using the first approach, but the upper limit on the temperature is 
significantly higher.  This second hypothesis can be plausibly
reconciled with collisoinal ionization predictions, so we consider it to be
the more likely of the two scenarios.  

\subsubsection{Q1626+6433: $z=2.321$ (Figure 12)}\label{sec_1626_2.32}

This unusual system is actually located 300 km/s {\em redward} of the \lya
emission line of the QSO, an effect that can be caused by a 300-500 
km/s blueshift of of the \lya emission line relative to the ``true'' 
QSO redshift as measured from narrow forbidden lines \citep{tytler1992}.  
However, the properties of this system favor its interpretation
as an intervening rather than ejected absorption system.  
Ultimately we do not include it in the cosmological sample 
because of its proximity to the background quasar. 

The \hi structure consists of a group of four
moderately strong lines ($14.0 < \log N_{\mhi} < 15.5$) within a $\Delta
v\sim 500$ km/s velocity interval - similar to what is seen in many of
the other \ovi absorbers we have detected.  \civ, \siiii, \siiv, and \ovi
are all seen in the vicinity of the central, strongest \hi line and
the velocity structure in all of these ions is identical except for 
\ovi.  The velocity profiles of the
moderate ionization lines are consistent with their production occurring
in a single common gas phase at $T=2.25\times 10^4$ K.

The \ovi profile is much stronger and broader than those of all the
other observed species.  A single $b_{\movi}=37$ km/s Voigt profile
provides our best fit to the data, and sets an upper bound on the 
temperature of $T_{\movi}\le 1.3\times 10^6$ K - almost two orders of
magnitude higher than the lower ionization species.

One could interpret the large velocity width of \ovi, and its high 
ionization state as a sign that the \ovi gas is ejected from the QSO
rather than intervening.  While this
possibility cannot be disproven, the 300 km/s redshift of the \ovi
line relative to \lya requires that the ejection velocity $v_{ej}\le
200$ km/s even after including a +500 km/s correction to the \lya
redshift.  This velocity is very small compared to values of several 
thousand km/s typical of material ejected from QSOs \citep{turnshek1984}.  
Furthermore, the close
alignment of the \ovi and other ions ($\Delta v < 40$ km/s), the
accuracy of the doublet ratio implying full coverage of the continuum
source, and the quiescent kinematics of \civ and
other lower ionization species,  all point away from the ejection
hypothesis for this absorber.  We favor the interpretation that
the absorption arises in intervening gas, possibly associated with the
QSO host galaxy or a nearby companion.

\subsubsection{Q1700+6416: $z=2.316$ (Figure 13)}\label{sec_1700_2.31}

This \ovi absorption is associated with a Lyman limit system, of column
density $\log(N_{\mhi})=17.6$, although only \lya and \lyb could be used 
in the measurement and both were saturated.  
Rich heavy element absorption is seen in both high
ionization species (\civ, \siiv, and \ovi) and
low ionization species (\cii, \alii, \siii).  

The low ionization species all share a similar kinematic structure,
concentrated around two lines separated by $\sim 45$ km/s in the very
core of the \hi profile.  The narrow widths of these lines imply low
temperatures in the range $T_{\movi}\le 0.7-7\times 10^4$ K, as would be 
expected in the central regions of a strong \hi system.  

The highly ionized species are very strong, and in
the case of \civ we observe significant saturation.
The velocity structure of the gas can be read from the \siiv line,
which appears to match the \civ profile but is probably not related to
\ovi.  The temperature of the C-Si phase derived from the \siiv 
line widths is quite cool, and resembles that of the low ionization
gas at $T\le 2-3 \times 10^4$ K.  However, the moderate ionization
species are much more widespread, spanning $\sim 150$ km/s.  
%
%
The \ovi profile is significantly contaminated by \lya forest
absorption, particularly in the 1037 \ang component. 
At least one line, located at $-20$ km/s in the Figure 1n, 
is clearly distinguished and may be associated
with the strongest \civ and \siiv lines even though the
redshifts do not align exactly.  
Our best-fit Voigt profile for this line contains two components,
with $b$ parameters of $14.35$ and $11.53$
km/s ($T_{\movi}=1.27-1.97 \times 10^5$ K) and column densities of
$\log(N_{\movi})=14.0, 14.2$.  The column densities derived from fitting
the rest of the profile are all similar, though the linewidths are 
larger ($b=15-25$ km/s).  While this may be an indication of higher
temperature gas, the model for this part of the \ovi profile is highly
uncertain because of the strong forest contamination.

\subsubsection{Q1700+6416: $z=2.379$ (Figure 14)}\label{sec_1700_2.37}

This is the most kinematically simple \ovi system we have detected in
our survey; each of the observed metal lines contains only a single
absorption component.  The \hi line is at the strong end of the
\lya forest regime, with two dominant components at $\log(N_{\mhi})=14.8,15.4$.  
Besides \ovi, we detect most of the common high ionization
species, including \civ, \siiii, and \siiv, however no \nv is
detected.  We also do not see any low ionization lines.

Our fits to the Si and C lines may be accurately explained with a
single gas phase, with $T_{C,Si}=3\times 10^4$ K, characteristic of low
density gas in the \lya forest.  The \ovi line differs from these
other species slightly in both redshift, which is offset by 10 km/s
(a $6\sigma$ difference), and doppler width, measured at 19 km/s.  
This width admits
the possibility of a high termperature gas, i.e. $T_{\movi}\le 3.4\times 10^5$
K.  Although one might expect some uncertainty in the line widths
because of blending in the \ovi 1032 line, we have used only the 1037
A component of the doublet in our fit, and found that it matches the
1032 \ang profile extremely well.  The measured column densiy ratios for
\ovi, and \siiv are $\log(N_{\movi}/N_{\mciv})=0.323$, and 
$\log(N_{\movi}/N_{\msiiv})=1.570$.  As was discussed in Section
2.5.12, these ratios do not agree with the predictions of simple
photoionization models for \civ and \siiv, so it is quite likely 
that the \ovi is physically distinct from these other species.  

\subsubsection{Q1700+6416: $z=2.436$ (Figure 15)}\label{sec_1700_2.43}

This complex system contains a number of lines distributed
over $\sim1000$ km/s.  The lines are grouped into two distinct clusters
separated by $\sim5-600$ km/s, and each cluster contains absorption from
both low ionization species (e.g. \cii) and high ionization species,
including \civ, \siiii, \siiv, and \ovi.  

The \hi absorption in this system
contains 8 lines with $\log(N_{\mhi})>14.0$ in a $1000$ km/s velocity
interval.  The \hi near the blue line cluster is dominated by a 
single line with $\log(N_{\mhi})=16.943$.  The red cluster's \hi
splits into several components in the higher order Lyman
lines, and is dominated by three lines with $\log(N_{\mhi})=14.860,
15.706$, and $15.806$.  The sharp line seen between the two strong \hi
features is an unrelated \siiii 1206 line at $z=2.463$.  

In both of the line clusters, blending from the \lya forest complicates the 
measurement of \ovi parameters.  This is apparent for the absorption at
$-300$ km/s in Figure 1p, for which the 1037\ang component is blended 
with a strong \hi line.
Though not obvious in the Figure, it is also true for the absorption 
at $+300$ km/s, which is blended with a $z=2.46$ \lyb line.  
By fitting the corresponding \lya line ($\log(N_{\mhi})=14.4$), we were
able to remove the \lyb contamination, revealing the profile shown in
the figure.
%
%

Kinematically complex \civ is seen in the neighborhood
of each of the absorption clumps, and \siiii and \siiv are also
strong, through not as distributed in velocity.  Low ionization
species (\cii and \siii) are
seen in the same vicinity as some \siiii and \siiv, though only near
the strongest of the \hi lines in the blue line cluster.  The
linewidths of the low and moderate ionization gas are
consistent with thermal broadening at $T=4-7\times 10^4$ K.  
%
%

The kinematics of the \ovi gas are more difficult to constrain because of
\lya forest contamination.  For the blue line cluster, we have treated
all the absorption in the \ovi 1032 \ang line as actual \ovi.  Under this
assumption, a three component model is sufficient to describe the
data.  Two of the components can be matched to the 1037 \ang profile in
the wing of the blended \lya, but the third 
component is completely blended and therefore
more suspect.  All three of the lines are broad; the doppler widths for 
the two secure lines are $b_{\movi}=22.14, 23.48$ km/s, or
$T_{\movi}\le 5.3\times 10^5$ K.  In the redder cluster of
lines, the velocity widths are difficult to measure because the
profile is not easily described by discrete lines.  Our best-fit model
again shows lines that are broader than those of other ions, 
near $b_{\movi}=15$ km/s or $T_{\movi}\le 2.2\times 10^5$
K.  One line is much broader yet at $40$ km/s.  

\subsubsection{Q1700+6416: $z=2.568$ (Figure 16)}\label{sec_1700_2.56}

This system has a simple velocity structure, and is
centered near a moderate \lya absorber ($\log N_{\mhi}=14.5$).  A
single, narrow heavy element line with weak flanking
absorption is seen in \civ and \siiii; a very weak \siiv feature is
also detected, but is not strong enough to provide detailed velocity
information.  
The \civ linewidth is an intermediate $b=12.9$ km/s ($T\le
1.2\times 10^5$), while \siiii is very narrow, implying $T\le 2\times
10^4$ K. 

The shapes of the \ovi 1032 and \ovi 1037 \ang components do
not match in detail, probably because of blending in the 1032 \ang line.
The \ovi 1037 line profile does however resemble the shape of  
\civ, although our best fit shows it to be offset by 10-12 km/s.  
Our estimate of 17.8 km/s for the \ovi linewidth implies a higher
temperature for the \ovi gas, with an upper limit of $T\le 3.0\times
10^5$ K.  However, we caution that blending in the \ovi core makes
it difficult to obtain a reliable measurement of the width.  While the 
coincidence with \civ increases the probability of this system being
\ovi, we still regard its identification as tentative.  

\subsubsection{Q1700+6416: $z=2.716$ (Figure 17)}\label{sec_1700_2.71}

The system is another example of gas ejected from the background
quasar, in this case at 2250 km/s.  The system was identified by
an unmistakable correspondence between the profiles of the \ovi doublet
components.  Strong absorption is seen in \ovi, \civ, and \nv, and in 
each case the ratio of the
line strengths for the two components of the doublet is smaller than
the expected value, indicating partial coverage of the continuum
source.  Because this is an ejected rather than intervening source, we
have excluded it from further analysis.

\subsubsection{Q1700+6416: $z=2.744$ (Figure 18)}\label{sec_1700_2.74}

The last system we discuss contains four distinct components, neighboring
a single weak \hi line.  No heavy elements other than \ovi are
detected.  The measured linewidths for the three \ovi components 
range from $7.14$ km/s to $15.27$ km/s ($T_{\movi}\le 5\times 10^4 -
2.2\times 10^5$ K.  Since this system is extremely close to the
background QSO, we do not include it in our cosmological statistics.
The quasar is known to be ejecting \ovi (see previous section), so
the absorption could be caused by additional outflow.

\section{Analysis}

\subsection{The Physical Environment of \ovi Absorbers}

The production of \ovi is thought to take place in two very different 
environments: either low density, photoionized
plasmas such as would be found in the \lya forest, or shock heated,
collisionally ionized plasmas, as would be found at the interface between 
dense structures and the general IGM.  The common association
we observe between \ovi absorbers and strong, metal-rich 
\hi systems causes us to favor {\em a priori} the second hypothesis.  
In the following sections, we draw comparisons between the measured
properties of the \ovi absorption and the predictions of 
ionization simulations to see if the two
are consistent with this qualitative conclusion.

For the ionization calculations, we have used the CLOUDY96 software
package \citep{ferland1998}.  The gas was modeled as an optically
thin plane-parallel slab in the presence of a \citet{haardtmadau1996}
shaped ionizing background spectrum for $z=2.5$.  The intensity of the
ionizing background was normalized to $J_{-21}=1.0$ \citep{scott2000}, 
and the gas
assumed to have a metallicity of $Z=\frac{1}{10}Z_\sun$ with solar
relative abundances for the heavy elements.  A grid of
models was then computed by varying the gas density and examining the
resulting ionization fractions and column densities of observable
ions.  For one of the runs, the temperature was allowed to converge on 
the thermal photoionization equilibrium value.  Then, subsequent runs were
performed at increasing fixed temperatures, to examine the effect of 
collisional processes on the ionization balance of the gas.  


\subsubsection{Pathlength and Gas Density Constraints}\label{sec_pathlength}

With the aid of the ionization simulations, it is straightforward
to calculate the absorption pathlength through a cloud given its
column density:
\begin{equation}
\label{eqn_pathlength}
L= {{N_{\movi}}\over{n_H f_{\movi}}} \left({{{O}\over{H}}}\right)^{-1}.
\end{equation}
Here the Oxygen abundance is an input to the simulations, and the
ionization fraction $f_{\movi} = n_{\movi}/n_{O}$ is output as a function
of the gas number density $n_H$.
Figure \ref{fig_pathlength} depicts this relation for a cloud with
$N_{\movi}=10^{13.5}$ \pcmsq~ of \ovi absorption, which
corresponds approximately to the weakest system
in the survey.  The right axis of the plot shows the instrinsic line
broadening expected for structures of different sizes due to Hubble 
expansion.  Again, we assume a flat, $\Omega_M=0.3,
\Omega_\Lambda=0.7$ cosmology, for which $H(z=2.5)=239h_{65}$ km s$^{-1}$ 
Mpc$^{-1}$.  The solid line represents the photoionization equilibrium
solution; other lines represent fixed temperature solutions as indicated.

By comparing the broadening due to the Hubble flow with the observed 
distribution of $b$ parameters, the data shown in Figure
\ref{fig_pathlength} may be used to constrain the
sizes and gas densities of the \ovi absorbing regions.
In the figure, we have shaded the region above $\Delta v_{\rm
Hubble}=48$ km/s, which represents the maximum $b$ parameter measured
from the \ovi lines in the survey.  A series of
tests has shown no substantial dropoff in the completeness of our sample
with increasing $b$ out to $b\sim50$ km/s, except for the weakest
lines (See Section \ref{sec_xsection}). 
We therefore expect that we could have detected modestly broader lines 
if they existed in the data, though it is possible that very broad lines
($b\sim100$ km/s) might be missed.  

Referring to Figure \ref{fig_pathlength}, the choice of 
$b_{max}=48$ km/s coupled
with a pure photoionization model provides the most conservative
lower limit on the \ovi gas density, at $n_H \ge 2\times 10^{-5}$ for a
cloud of size $L\le 200$ kpc (shown as the unshaded region).  
According to this prescription, and assuming that $\Omega_b
h_{100}^2=0.02$ \citep{omeara2001}  we
find that at $z=2.5$ the \ovi absorption lines arise in structures
with $\rho/\bar{\rho}\ge 2.5$.  
In reality, most of the lines we observe are much
narrower, with a median width of $b_{\movi}=16$ km/s.  Use of this value
instead of the more conservative $48$ km/s yields a 
characteristic pathlength of $L\sim 60$ kpc, and a density limit
roughly twice as high.  
If the model metallicity were reduced to $0.01Z_\sun$, the lower limit 
on $\rho/\bar{\rho}$ would rise further by a factor of $\sim 3$, or
if the gas were hot and collisionally ionized $\rho/\bar{\rho}$ 
would increase by a factor of $\sim 3.5$.  

In short, we find from comparison of the simulations to observed
linewidths that \ovi absorbers have sizes of $L\le 200$ kpc and
overdensities of $\rho/\bar{\rho}\ge 2.5$, with true typical values
probably nearer $L\sim 60$ kpc and $\rho/\bar{\rho}\sim 10-30$.  
Cosmological simulations suggest that such structures may correspond to
previously metal-enriched 
gas which is in transition from the cool, distributed 
\lya forest to a denser, more compact phase 
\citep{zhang1995,cen1994,censimcoe1997}.  
It is in this
neighborhood ($10^{-5}<n_H<10^{-4}$) that the temperature-density 
relation for the photoionized \lya forest begins to break down due to 
shock heating of the infalling gas \citep{mcdonald2001}.

\subsubsection{Temperature Structure}\label{sec_temperature}

In principle, one of the most powerful methods for distinguishing
between photoionized and collisionally ionized \ovi is simply to
measure the temperature of the gas.  Clouds found in the
more tenuous regions of the IGM are in 
photoionization equilibrium, and exhibit a maximum
characteristic temperature of $T\sim 40,000$ K.  The effect of 
collisional ionization at this temperature
is minimal.  Indeed, collisional processes only become important above 
$T>10^5$ K, and as the temperature rises they soon dominate the
physics of the gas regardless of its density.  The \ovi
ionization fraction itself peaks at $T\approx 10^{5.5}$ K 
\citep{sutherland1993}.  

In practice this ideal, bimodal distribution of \ovi temperatures
is not observed, because line broadening in low density 
photoionized gas is enhanced by Hubble expansion and peculiar velocities.
However, we find that the statistical distribution of \ovi linewidths 
relative to those of other species does provide some evidence that the
\ovi gas is distinct from that of the other ions.  
Figure \ref{fig_b_temp} illustrates these distributions for \civ,
\siiv, and \ovi, expressed as the temperature $T_{max} =
A(b/0.129)^2$, where $b$ is the measured line width and $A$ is the
atomic mass number of each ion.
Since this estimate of the temperature does not account for 
non-thermal broadening, $T_{max}$ represents only an 
upper bound on the gas temperature for a given line.  

Despite this limitation, we find that the \ovi distribution
differs significantly from those of the other ions in the sense that it
favors broader lines.  A qualitative similarity between the
\siiv and \civ distributions is confirmed by a Kolmogorov-Smirnov
test, which indicates with $88\%$ probability that they are
drawn from the same parent population.  The probablility that the \ovi 
shares the same parent distribution was found to be only $0.0006\%$.  
The evidence suggests that the gas producing the \ovi absorption is 
physically distinct from, and probably hotter than a
second phase which is the source of both \civ and \siiv.  While the
temperatures plotted represent only upper limits, it is intriguing
that the \ovi histogram is distributed roughly evenly about
$\log(T_{\rm max})=5.5$, the temperature at which the collisional ionization of
\ovi peaks (shown as a solid vertical line).  In fact, the median
value of $T_{\rm max}$ for \ovi is $2.1\times 10^5$ K, and 
and 62\% of all systems lie in the range $5.0 \le \log T_{max} \le 6.0$.
While this does not constitute direct evidence of a hotter,
\ovi phase with significant collisional ionization, it is at least 
consistent with such an interpretation.  

\subsubsection{Multiphase Structure}\label{sec_ionization}

The evidence presented thus far indicates that high redshift \ovi is 
found in the neighborhood of complex systems where multiple  
gas phases coexist.  In this section, we examine the 
CLOUDY predictions of relative absorption strengths for different ions, in 
an attempt to identify the distinguishing physical characteristics of 
these phases.  We recall that for these calculations, the relative
element abundances are held fixed at the solar level.  

Figures \ref{fig_civ_siiv} and \ref{fig_ovi_civ} illustrate the 
predictions for the column density ratios \civ/\siiv and \ovi/\civ 
(expressed in logarithmic units).  In both plots, the solid line 
represents a thermal photoionization equilibrium (PIE) solution,
and other lines represent runs made at fixed temperatures
to isolate the effects of collisional ionization as the gas is heated. 
The PIE curve and the $T=10^{4.5}$ K curve are quite similar for both
\ovi/\civ and \civ/\siiv, indicating as expected that collisional 
processes have minimal effect on the ionization balance at low 
temperature.  As the temperature rises, collisional
processes tend to drive the column density ratios toward the 
expected value for pure collisional ionization equilibrium
($2.1$ for \ovi/\civ, $1.9$ for \civ/\siiv, again in logarithmic units).  
This value is roughly density-independant, as the collisional 
ionization rate and recombination rate both vary as $n_H^2$.  
Accordingly, at high densities where PIE tends to favor
lower ion ratios than does CIE, the ratios rise with temperature.
In both cases this transition is quite abrupt - we see that the high
density gas changes from a photoionization-dominated state to
a collisionally ionized state over a factor of 3 range in 
temperature.  At low densities, the tradeoff between collisional and
photon processes differ in Figures \ref{fig_civ_siiv} and 
\ref{fig_ovi_civ}.  For \civ/\siiv, at low density the CIE
ratio is actually lower than the PIE value, so an
increase in temperature only increases the
recombination rate, and drives the ratio downward.
For the \ovi/\civ case, the CIE and photoionization equilibrium ratios
are similar, so the combined processes enhance the \ovi/\civ
ratio 7-10 times higher than the expected value for either process
acting alone.  

In Figure \ref{fig_civ_siiv}, we have shaded a region which
approximates the $\pm 1\sigma$ range of values for \civ/\siiv that we
detect in the survey.  The only region where the data and models
overlap is at high density and low temperature: the data rule out 
models with densities above $n_H\sim 3\times 10^{-3}$, or 
gas temperatures above $10^5$ K at any density.  Evidently, the 
\civ and \siiv absorption arise in a
fairly dense phase ($n_H\sim 5\times 10^{-3}$) which is in 
photoionizaton equilibrium with the UV background, at 
$T\sim 1-5\times 10^4$ K.  Such a population is essentially identical
to the \civ and \siiv systems commonly observed in all quasar spectra, 
and has been extensively characterized \citep{songaila1996}.  

In Figure \ref{fig_ovi_civ}, we see that for the densities
and temperatures inferred for \civ and \siiv, the predicted \ovi 
strength should be quite small - 10-100 times weaker than \civ, 
and therefore likely
undetectable.  More generally, the ionization simulations imply that
{\em whenever} coincident absorption of \civ and \siiv is observed, 
models require a separate, colder, condensed gas phase that is
unlikely to be seen in \ovi.  This argument may be extended to other
low ionization species such as \siii, \siiii, or \cii as well.  
In other words, any strong \ovi observed in connection with these systems 
{\em cannot} come from the same gas as the
\civ and \siiv - it must be contained in a separate gas phase.  This
conclusion is consistent with the interpretation of the
temperature distributions given in \ref{sec_temperature}.
It is also consistent with the visual appearance of the
line profiles, whose C and Si substructures match closely, and 
appear to be independant of the \ovi substructure.

It is difficult to place firm constraints on the ionization mechanism 
for \ovi itself precisely because of
this multiphase nature.  In the \civ region, relatively
strong and defined absorption from the cool Carbon-Silicon phase masks the
presence of any weak \civ that might be associated with a hot \ovi
phase, rendering an accurate measurement of \ovi/\civ impossible.  
In theory, the \ovi/\nv ratio should also provide a useful ionization
probe, but unfortunately in our spectra the \nv region is
contaminated by interloping \lya forest absorption 
in nearly every \ovi system.

For two systems, it is possible to obtain accurate lower limits 
on the \ovi/\civ ratio due to velocity offsets between \ovi and the cold
\civ /\siiv phase.  In Figure \ref{fig_ovi_civ}, 
we have shown these measurements as horizontal lines with arrows.  We
have extended the lines over the range in density framed by the complete
photoionization model and the complete collisional ionization model,
except where this violates the shaded constraints which are described
below.  For one of these systems (located at
$\log(N_\movi/N_\mciv)\sim 2.5$) the lower limits cannot be explained
except through collisional ionization.  For the other (located at
$\log(N_\movi/N_\mciv)\sim 1.4$ due to lower signal-to-noise in the \civ
region) a pure photoionization model is allowed, but only at $n_H\sim
10^{-4}$.  This is significantly lower than what is seen for the cooler
\civ/\siiv phase present in many absorbers.  Collisional solutions are
also allowed for this system over a wide range of densities.  

Given that the accuracy of direct \ovi/\civ measurements is
limited, we instead examine Figure \ref{fig_ovi_civ} to determine
which regions may be reasonably excluded from consideration.
First, based on the pathlength/density constraints described in 
\ref{sec_pathlength}, we exclude the shaded region to the right of the
figure corresponding to the $\rho/\bar{\rho}\sim 1-2.5$ \lya 
forest.  The lower left region of the diagram, corresponding to a high
density phase in PIE, may also be ruled out because it cannot be
simultaneously reconciled with the \civ and \siiv line strengths.

Two allowed regions remain: one to the upper left of the diagram,
and one covering the middle of the plot, in the density range 
$2\times 10^{-5}<n_H<2\times 10^{-4}$.  The first of these corresponds
to condensed structures with similar overdensities as the C-Si phase
($\rho/\bar{\rho}\ge 100$), but which are hot ($T>10^5$ K) and
essentially entirely collisionally ionized.  The second allowed region
covers the overdensity range $3\le \rho/\bar{\rho} \le
30$, in the transition area between the \lya forest and condensed
structures.  Both photoionization and collisional ionization are viable
mechanisms for \ovi production in this regime.  Since the density is
only an order of magnitude higher than 
in the \lya forest the gas can be efficiently ionized by the UV 
background, but simulations also indicate that as structures with
these densities collapse, shocks at the IGM interface heat the gas 
to $T=10^5-10^7$ K where collisional ionization should dominate 
\citep{cen1999,fang2001}.  
It is likely that the ionization balance in this density range is 
actually governed by a mixture of the two processes, where gas that 
is already somewhat highly photoionized has its ionization level
enhanced further upon rapid heating.   

The results of this section may be summarized as follows.  Comparison
of \ovi, \civ, and \siiv line ratios indicates that a multiphase
structure is required to explain the simultaneous existence of highly
ionized \ovi and lower ionization species such as \siiv.  The low 
ionization elements are contained along with \civ in a cool, condensed 
phase with overdensities of $\rho/\bar{\rho}>100$ and 
$T\sim 1-5\times 10^4$ K.  This gas is in photoionization equilibrium 
with the UV background and its properties are identical to the \civ
and \siiv systems commonly seen in quasar
spectra.  The \ovi gas is contained in a separate phase, which traces
either high density ($\rho/\bar{\rho}>100$), high temperature ($T>10^5$ K) 
gas, or structures of overdensity $3 < \rho/\bar{\rho}<30$ at the
boundary between condensed regions and the distributed IGM.

\subsubsection{Constraints on Number Density and Cross-Section}\label{sec_xsection}

Let us now examine constraints on the number
density ($n$) and cross section ($\sigma$) of high redshift \ovi 
systems.  In this context, we refer to a ``system'' as a physical
complex of gas, which may contain several individual subcomponents or lines.
Assuming that both $n$ and $\sigma$ are constant throughout the 
survey volume, the expected number of \ovi detections is given as: 
\begin{equation}
\label{eqn_nsigma}
N = f_c \left({n\sigma \times \frac{c}{H_0} \sum \Delta X}\right),
\end{equation}
where $\sum \Delta X$ is the survey pathlength
defined in Equation \ref{eqn_pathlength}.  The factor $f_c$ is 
an estimate of the survey's completeness, and is included to account
for lines that remain undetected due to blending from the \lya forest.

To quantify the completeness, we have run a series of tests where
artificial \ovi doublets were added to each survey sightline to
measure the recovery efficiency of our seach method.  The column
densities and doppler $b$ parameters of these lines were assigned 
according to the same
joint probability distribution $p(b,N)$ as the actual lines detected
in the survey.  Furthermore, we imposed a lower limit on the \ovi
column density of $\log(N_{\movi})\ge 13.4$ - the strength of
the weakest intergalactic system found in the survey.  Five such realizations
were generated for each quasar sightline, and the numbers and
properties of the artificial lines were hidden from the users to avoid 
any bias in the search process.  Such a sample
does not test whether the systems detected in the survey are
representative of the parent population of all \ovi absorbers;
however, it does provide an accurate estimate of the fraction of
systems like those observed that remain undetected.

Upon searching the generated spectra in the same manner as the survey data, 
we consistently recovered $\sim 40\%$ of the artificial \ovi lines
with individual sightlines ranging from $38\%$ to $44\%$.  
Closer examination revealed that essentially all of the unrecovered lines
were lost due to blending with \hi rather than a poor signal-to-noise
ratio.  In clean portions of the spectra, our detection threshold
varied between $12.0\le\log(N_{\movi})\le13.0$, which is almost an
order of magnitude below the level where blending from the forest
begins to cause completeness problems.  Generally our search method was
most effective at recovering systems with $\log(N_{\movi})\ge 13.5$.
Above this column density the completeness is probably
$45-50\%$ and shows surprisingly little dependence on the doppler
parameter until one reaches $b\le 10-15$ km/s, at which point the
efficiency goes up dramatically.  For $b\ge 15$, the
completeness is nearly constant with $b$ at $\sim 35\%$ all the way to
$b\sim 50$, the largest doppler parameter measured in the survey.  
At column densities below $\log(N_{\movi})=13.5$ the completeness
begins to drop, and not surprisingly the lines that are lost
tend to be broad, weak lines in blends.  However, even at lower column density
we recover the narrow lines ($b<20$ km/s) in roughly $20-25\%$ of the
cases.  For the calculations below, we have adopted a completeness
level of $f_c=0.41$, which corresponds to the average for all
sightlines, and includes the entire range of column densities and
doppler parameters seen in the survey.

Given that we have detected 12 intergalactic systems in a path of $\sum
\Delta X = 6.90$, we can invert Equation \ref{eqn_nsigma} to constrain the
product of the space density and size of the \ovi absorbers.  For 
spherical \ovi clouds with (proper) radius $R$ and (proper) density
$n$, we find
\begin{equation}
\label{eqn_nsig_norm}
\left({{n}\over{{\rm 1 ~Mpc^{-3}}}}\right)\left({{R}\over{{\rm 17
 ~kpc}}}\right)^2 = 1.0.
\end{equation}
We note that a characteristic cloud size of $L=2R\sim40$ kpc
is roughly consistent with estimates made using the completely
different method of pathlength calculation from ionization simulations
described in \ref{sec_pathlength}.
The implied volume filling factor for such clouds is quite small at $f_v=
2.0\times 10^{-5}\left({{R}\over{18 {\rm
kpc}}}\right)$.  Taking into account the size
constraint $R=L/2<100$ kpc from Section \ref{sec_pathlength}, 
we estimate $f_v \le 1.1\times 10^{-4}$. 
It therefore seems unlikely that the
clouds are generally distributed, as in the \lya forest whose filling
factor is much more substantial.  Rather, the cross sectional
constraint is more consistent with a compact topology, as one might
expect to find in the surrounding areas of galaxies or clusters at the
intersection of more extended, filamentary structures.

\subsubsection{Clustering Behavior}
In Figure \ref{fig_tpcf}, we show the observed two-point correlation 
function (TPCF) of the \ovi absorbers.  Special care was taken 
to account for the effects of spectral blockage from the \lya forest
when calculating the TPCF.  
To produce the version shown in the figure, we have simulated 1000
realizations of the set of 5 sightlines from the survey.  Selection functions
were carefully constructed for each sightline, excluding regions 
where strong lines in the actual data prohibit \ovi detection (roughly 
$\tau > 0.5$).  Each simulated sightline was populated
with \ovi at random redshifts, requiring that 
both components of the doublet fall in an unblocked region of the 
spectrum.  The randomly distributed
systems were then collated by pairwise velocity separation, and
compared to the distribution of velocity separations in the data to
calculate the correlation: 
$\xi(\Delta v) = N_{{\rm data}}(\Delta v) / N_{{\rm random}}(\Delta v)
- 1$.  

Although we have only identified 12 \ovi systems in the data, each
system has several subcomponents and for the calculation of the 
correlation function we have used the full set of these subcomponents
as determined by VPFIT.  The TPCF constructed in this way
probes the structure both within individual systems and 
between different systems.  Poisson errorbars ($1\sigma$) are shown 
in the Figure to indicate the level of uncertainty due to finite sample size.  

The data show evidence for strong clustering at separations of $\Delta
v < 300$ km/s, and a weaker signal out to $\Delta v \sim 750$ km/s.  
This clustering pattern differs from that of the \lya forest, which 
shows almost no signal except for a weak amplitude at the shortest 
velocity scales ($\Delta v \sim 100$ km/s, Cristiani et al 1997).
We have also calculated the TPCF for the \civ absorption associated
with our \ovi selected systems, using the same window function, to
test for differences between the \ovi and \civ clustering properties
within our sample.  Except at the smallest velocity separations
($\Delta v < 100$ km/s) which probably reflect the internal dynamics
of the systems more than their spatial clustering, we find no statistically
significant differences between the \ovi and \civ clustering in the
\ovi selected systems.  At small
separations the \ovi clustering amplitude is $\sim 50\%$ lower than
that of \civ, likely a reflection of the smoothness of the \ovi 
absorption profiles.

For reference, the Figure \ref{fig_tpcf} also shows the TPCF for 
a larger sample of \civ absorbers that is not \ovi selected, 
taken from \citet{rauch1996}. In general, the \ovi correlation 
resembles that of the larger \civ sample.  There is some evidence for
an enhancement of \ovi clustering at the $\sim2\sigma$ level for 
$70\le\Delta v \le 200$ km/s.  This excess clustering is seen in
the \ovi selected \civ lines as well, and may reflect the qualitative
impression that \ovi is often associated with very strong,
kinematically complex absorption systems.  
More inclusive samples such as \citet{rauch1996} pick up a larger
fraction of weaker \hi systems that have one or a few \civ components.

The range over which we measure correlation signal in \ovi
extends to larger values than the $100-200$ km/s 
typical of virialized, galactic scale structures.  This could mean that 
the individual \ovi lines, which arise in structures of $L\sim 60h_{65}^{-1}$ 
kpc, actually trace large scale structure in an ensemble
sense.  According to this picture, the signal at large velocity 
separations - corresponding to coherence at $L\sim 7-10h_{65}^{-1}$
comoving Mpc scales - 
results from chance alignments with cosmological filaments along the
line of sight.  The solid line in Figure \ref{fig_tpcf} represents a
simple power law fit to the TPCF over the range $100\le \Delta v\le 1000$
km/s, where the effects of peculiar motions should be less severe.
We find:
\begin{equation}
\label{eqn_powerlaw}
\xi(\Delta v) = \left({{\Delta v}\over{750~ {\rm km/s}}}\right)^{-1.68}.
\end{equation}
%
%

The form of this power law is similar to its three dimensional analog
measured in local galaxy redshift surveys, which find best fit exponents of 
$-1.7<\gamma<-1.8$ \citep{connolly2001,nordberg2001,LeFevre1996,loveday1995}.
For comparison, we have shown the correlation function of local galaxies
measured from the APM survey \citep{loveday1995} as a dashed line in
Figure \ref{fig_tpcf}.  If the high velocity tail of the \ovi
TPCF is driven by cosmological expansion alone, then the correlation 
length for \ovi at $z=2.5$ is $L=10.8h_{65}^{-1}$ comoving Mpc.  
This is slightly larger than the galaxy correlation length found in the
APM and other local surveys, but it is less than that of clusters at the
present epoch \citep{bahcall1988}.  At higher redshift, Lyman break 
galaxies show a similar power law slope, though the
amplitude is somewhat weaker than for \ovi \citep{giavalisco1998}.  
In most structure formation models the the overall normalization of
the power law is predicted to evolve with redshift \citep{benson2001}
but not the slope.  The -1.7 power law for the \ovi TPCF is therefore
a natural outcome if the \ovi systems are tracing large scale
structure. 

Thus far we have ignored the effects of peculiar
velocities on the TPCF.  The signature of gas motions within the
potential wells of $L^*$ type galaxies has been well documented
\citep{sargent1988}, and it is thought that such motions are 
unlikely to account
for signals on scales of $\Delta v > 150$ km/s.  However, for highly
ionized gas such as \ovi one might expect to see correlations on larger
velocity scales caused by galactic winds.  In this scenario, pairwise velocity 
separations of $400-800$ km/s are produced in bidirectional outflows which 
drive material both towards and away from the Earth at projected
velocities of $200-400$ km/s.  

To characterize this effect on the TPCF, we consider a naive, spherically
symmetric outflow model where all material is ejected from a central
source at a common velocity $v_0$.  Assuming the outflow is finite in 
size, the area over which a given velocity splitting $\Delta v$ can be 
observed scales as 
\begin{equation}
A(\Delta v)\propto 1 - \left({\Delta v}\over{2 v_0}\right)^2, 
\end{equation}
with a maximum observable $\Delta v = 2 v_0$ when the outflow vector
is parallel to the line of sight.
The area-weighted mean velocity splitting for this model is
$\left<{\Delta v}\right>=\frac{3}{4}v_0$.  For $v_0=400$ km/s, which
is chosen to match the high-end cutoff in the TPCF for our data and
represents a reasonable ejection velocity for starburst-driven winds 
\citep{lehnert1996,pettini2001},
the model predicts a maximum in the correlation signal 
at $\Delta v \sim 300$ km/s.  It also predicts that
roughly equal numbers of pairs should be observed with splittings
above and below $300$ km/s - i.e. the TPCF should be much flatter than
what is actually observed.  This problem with the wind model could
in principle be solved by assuming a distribution in wind ejection
velocities skewed toward low values, or a hybrid model where the
correlation signal at low $\Delta v$ is dominated by motions within
galaxy potentials and the large separation pairs are caused by winds.
Given that such models would require significant tuning, and that a
power law fit motivated by local observations of large scale structure
matches the data reasonably well, the simplest interpretation is that the
\ovi TPCF signal is dominated by large scale structure.

Ultimately, comparison of the \ovi TPCF with simulations of structure
formation can provide a more realistic physical description
than we attempt here.  Also, we note that the lack of strong
outflow signature in the \ovi-\ovi TPCF does not preclude the
existence of \ovi rich winds.  In a bipolar outflow geometry, a
single sightline might only encounter one of two lobes.  In this case,
velocity offsets between \ovi and other ions such as \civ provide
better probes of the wind physics than \ovi-\ovi comparisons.

\subsection{The Contribution of Warm-Hot Gas to $\Omega_b$}\label{sec_omegab}

Having completed our analysis at the level of individual systems, we
now estimate the contribution of the total ensemble of \ovi systems to
the baryon budget.  For this discussion, we shall use the term 
$\Omega_{\rm WH}$ to denote the cosmological mass density of  
the {\em complete warm-hot gas mixture} containing \ovi; when referring only
to the contribution of quintuply ionized Oxygen atoms to the closure
density we use the term $\Omega_{\movi}$.  The former contains
ionization and abundance corrections which are highly uncertain, while
the latter is a direct observable.  For the calculation of 
$\Omega_{\rm WH}$, we use a slightly modified version of the the 
standard formula \citep{tripp2001,burlestytler1996} for the
cosmological mass density:
\begin{equation}
\label{eqn_omega}
\Omega_{\rm WH} = \frac{1}{\rho_c} \times {{\mu m_{H}}\over{f_{\movi}}}\left({\frac{O}{H}}\right)^{-1}\times{{\frac{1}{f_c}\sum N_{\movi,i}}\over{\frac{c}{H_0}\sum \Delta X_i}}.
\end{equation}

The outer terms in this equation are easily calculated,
the first being a normalization to the critical density at the present
epoch, and the third being the cosmological number density of \ovi ions,
expressed as the ratio of the total \ovi column density to the total comoving
pathlength of the survey.  The $\frac{1}{f_c}$ factor accounts for
incompleteness, and its determination is described in Section
\ref{sec_xsection}.  The second term - a conversion from number
density of \ovi ions to mass density of a gas mixture containing \ovi
- is highly uncertain and contains the abundance and
ionization fraction corrections mentioned above.  
For ease of comparison with local \ovi surveys, we have adopted
the same values as \citet{tripp2000a} for these quantities - a mean 
atomic weight of
$\mu=1.3$, an \ovi ionization fraction of $f_{\movi}=n_{\movi}/n_O=0.2$,
and an Oxygen abundance corresponding to $\frac{1}{2}Z_\sun$ for 
$\log\left({\frac{O}{H}}\right)_\sun=-3.17$ \citep{grevesse1998}.
As noted by Tripp et al, the values for the ionization fraction and
Oxygen abundance represent upper limits - indeed, at high
redshift the abundance in particular is likely to be lower by as much
as a factor of $50-100$.  As such, the value of $\Omega_{\rm WH}$ we derive is 
only a lower bound.

Combining Equation \ref{eqn_omega} with our survey results, 
the assumptions described above, and the incompleteness correction, 
we obtain the following lower limit on the cosmological mass density
of warm-hot gas: 
\begin{equation}
\Omega_{\rm WH}h_{65} \ge 0.00032,
\end{equation}
or about $0.5\%$ of the total $\Omega_b$.  This result is consistent 
with the generally accepted model of the high redshift IGM, where
most of the baryons ($\ge 90\%$) are contained in the relatively cool 
and less dense ($T\sim 10^4$ K, $\rho/\bar{\rho}\sim 1-5$) \lya forest 
network \citep{rauch1997b,weinberg1997}.  

Our result is very similar to the most
recent low redshift estimates of $\Omega_{\rm WH}$ from FUSE and 
STIS/HST,
which indicate $\Omega_{\rm WH}h_{65} \ge 0.00046$ at $z\sim 0$ - only 
$\sim 25\%$ higher than our $z=2.5$ estimate \citep{savage2002}
\footnote[2]{Note that for this calculation we have matched our 
  metallicity and ionization assumptions to the {\it most conservative} set 
  of assumptions from \citet{savage2002} and \citet{tripp2000a}.  The more
  commonly quoted value of $\Omega_{\rm WH}h_{65} \ge 0.002$ at low
  redshift is obtained by changing the assumed metallicity, so
  a comparison of this number with our survey would require a similar 
  rescaling of $\Omega_{\rm WH}(z=2.5)$ upward by a factor of 5.
  The $25\%$ evolution we quote is a change in directly 
  observable quantities (i.e. the third factor in Equation \ref{eqn_omega}).}.
It is also in agreement with the estimate of \citet{burlestytler1996}
made at intermediate redshift ($0.5<z<2$) using lower resolution FOS
data, provided one rescales their measured quantities to match our 
(and Tripp's) assumptions.

The weak evolution we observe in $\Omega_{\rm WH}$ is in
apparent contradiction with simulations,
which describe the properties of low redshift \ovi quite accurately,
but predict a decrease in $\Omega_{\rm WH}/\Omega_b$ from
$\sim 30-40\%$ at $z=0$ to $\sim 1-10\%$ at $z=2.5$ 
\citep{cen1999,dave2001,fang2001,chen2002}.  
Both our measurements and those of \citet{savage2002} are only lower limits,
but Equation \ref{eqn_omega} illustrates that with the observed
quantities essentially constant, the only way to produce a strong decrease in 
$\Omega_{\rm WH}$ towards higher redshift is to let the product of 
$f_{\movi} \times \left(\frac{O}{H}\right)$ {\it increase} with redshift.  
This is somewhat surprising since one generally expects the 
cosmic metallicity to be lower at earlier epochs.  

A possible explanation for this effect could be that the type of
system being traced by \ovi differs at high and low redshift.  For example,
the low redshift \ovi could be found in cosmological filaments and
sample metallicies closer to the cosmic mean, whereas the high
redshift systems could be located near galaxies or groups and be
subject to local metallicity enhancements and/or heating processes.  
Such a scenario would explain both the agreement between low redshift 
observations and simulations, and also the similar values of 
$\Omega_{\rm WH}$ at low and high redshift.

Another factor is that the \ovi we see
in absorption does not trace the bulk of the mass in the
warm-hot medium.  In fact, from Figure 5 of \citet{dave2001}, we see
that most of the gas - particularly at low redshift - should be hotter
than the optimal range for \ovi production, at $T>10^6$ K.
Prospects of detecting this gas are better in soft X-ray emission, or
absorption from higher ionization lines such as \ovii or \oviii.
Indeed spectra taken recently with Chandra-HETG may have already
revealed the presence of intergalactic $T>10^6$ K gas through such
absorption \citep{fang2002,nicastro2002}.

If the majority of the warm-hot intergalactic medium resides in a
reservoir of higher temperature gas, the relatively weak evolution we
observe in $\Omega_{\rm WH}$ might be explained if the \ovi traces a 
short-lived 
phase in the cooling cycle of a pre-heated plasma.  Recent cooling
models for highly ionized plasmas 
\citep{benjamin2001} show a peak in the cooling curve over the range
$10^5< T< 10^6$, where \ovi should be most abundant.  The interpretation
of evolutionary trends in $\Omega_{\rm WH}$ may be complicated
if the \ovi only measures an instantaneous snapshot of 
rapidly cooling non-equilibrium gas, 
rather than the total amount of shocked gas integrated over cosmic time.  

\subsubsection{Metallicity Constraints}\label{sec_metallicity}

Because of the multiphase nature of the \ovi systems, it is
impossible to directly measure the Oxygen abundance for individual
absorbers.  
Typically, the cool phase that gives rise to \civ and \siiv
also produces strong \hi absorption which overwhelms the
weaker \hi signal from the hot gas, and leads to erroneously small
estimates of [O/H].  Rather than attempting to
disentangle the phase structure of these systems, we estimate the
average metallicity of the gas in all \ovi systems, using reasonable
assumptions about the distribution of baryons in the high redshift universe.

Namely, if one assumes that the \ovi systems we detect are 
distinct from the \lya
forest, and that $\Omega_{{\rm Ly}\alpha}/\Omega_b \ge 0.9$ 
(i.e. $\Omega_{\rm WH}/\Omega_b\le 0.1$; Rauch et al 1997b, Weinberg
et al 1997), the
solution of Equation \ref{eqn_omega} may be inverted to place
constraints on the metallicity of the \ovi absorbers:
\begin{equation}
f_{\movi}\times\left({\frac{O}{H}}\right)\ge 
\frac{\mu m_H}{0.1~\Omega_b~\rho_c}~
{{\frac{1}{f_c}\sum N_{\movi,i}}\over{\frac{c}{H_0}\sum \Delta X_i}}=
4.3\times 10^{-6}
\end{equation}
Combining this with our assumed upper limit of $f_{\movi}\sim 0.2$, 
the corresponding limit on the metallicity is $[O/H] \ge -1.49$, or
$Z \ge 0.03Z_\sun$.  This limit is higher than the commonly quoted
value of $[O/H]\sim -2.5$ for the generally distributed IGM 
\citep{songaila1996,ellison2000,schaye2000}, but it is close to the average
abundance measured in damped \lya systems, which also exhibit
only minimal evolution in number density and metallicity over a wide 
range of redshifts \citep{prochaska2000,ellison2001}.  If correct,
this high metallicity estimate suggests that the \ovi systems in
our survey are found near regions with significant local enrichment
and do not probe the metal content of the more widespread IGM.

\section{Discussion}\label{sec_discussion}

In the preceeding sections, we have characterized \ovi systems by
their association with rare, high density environments in
the early universe.  However, we have not explicitly linked \ovi
production with any specific process occurring in this environment.
In this section, we examine two of the most likely
sources for \ovi production: shock heating of the 
IGM as it falls onto dense structures (hereafter referred to as 
the infall hypothesis), and the ejection of galactic superwinds
associated with high redshift star formation (hereafter the 
outflow hypothesis).  Both of these processes play an 
important role in the assembly of galaxies and the chemical enrichment
of the IGM, and both are known to occur at the redshifts
probed by our survey.  We cannot definitively distinguish which of
these dominates on the basis of the \ovi data alone, so we
weigh the merits of each below.  

\subsection{\ovi: Shocked Infall on Large Scale Structure?}\label{sec_infall}

There is currently little observational evidence either for or against
the existence of large-scale gas inflows at high redshift, but clearly
the process is required to take place at some time in any viable
structure formation model. In particular, the 
notion of \ovi as a tracer of a shock-heated,
warm-hot phase of the IGM has generated recent enthusiasm, 
because warm-hot gas provides a natural source of ``missing baryons''
at low redshift, and simulations have been fairly succesful at
reproducing the observed numbers and strengths of low redshift \ovi lines
\citep{cen2001,dave2001,fang2001,tripp2000b}.  In 
this scenario, low density gas in the IGM which has been chemically
pre-enriched crosses shock boundaries as it settles onto large-scale 
structure.  For post-shock temperatrues of
$10^5-10^7$ K, Hydrogen \lya absorption is suppressed by collisional
ionization and the production of \ovi and other highly ionized species
is enhanced.  At $z=0$, the simulations predict that this \ovi should
be visible in absorption, and that most of the absorption should occur
in widely distributed filamentary structure, away from the sites of
formed galaxies.

While the prediction of a {\em filamentary} warm-hot phase 
at first seems to be at odds with the analysis presented here, 
it is important to note that this was made for the
local universe, and that the warm-hot gas is expected to evolve
strongly with time.  \citet{dave2001}
have examined some of the evolutionary properties of the WHIM, and
demonstrated that as one moves to higher redshift, the production 
of warm-hot gas is governed by the amount of time structure has had 
to accumulate and shock heat infalling gas.  
We therefore expect that at high redshift \ovi should
trace more overdense structures than it does locally, since these
regions should be the first to accrete significantly and form shocks
strong enough to heat gas above the thermal photoionization
equilibrium temperature.  

The properties of the warm-hot gas 
whose evolution is explicity tracked by Dav\'e et al
seem to be in reasonable agreement with our observations at high redshift.
For example, at $2<z<3$, the simulations predict that the mass 
of the Warm-Hot medium is dominated by structures 
with $10<\rho/\bar{\rho}<30$, which coincides with our crude overdensity
estimates of $3<\rho/\bar{\rho}<30$ for \ovi systems.  
Furthermore, the authors claim that
warm-hot gas is distributed across a wide range in temperature, peaking 
near $10^6$ K with negligible effects from radiative cooling.  If most
of the gas is trapped in this hot state which is unable to radiatively
dissipate the heat generated from shocks, then as time progresses,
an increasing portion of the mass in the warm hot phase will pile up
in this reservoir that is too hot to be seen in \ovi.  Essentially,
all the gas that is heated to temperatures above the \ovi range will
remain in this hot state, while the gas which is heated only to \ovi
temperatures will cool very efficiently via line radiation until the
\ovi recombines and is no longer seen.  This could provide an
explanation for the similar values we observe for $\Omega_{\rm WH}$ at
low and high redshift, as the \ovi would effectively trace only the
instantaneous amount of gas undergoing shock heating, rather than the
total amount of gas which has passed through a shock at some time in
its history.
%

The overdensities seen in our survey fall in the 
range where structures in cosmological simulations make a topological 
transition from connected filaments to isolated spheres.  
Because of computational limits
it is difficult to accurately predict the both the numbers and sizes of these
clouds: the large simulation cubes required to minimize shot noise in
$n$ tend to resolve the structures of interest only marginally,
while higher resolution grids tend to sample a small number of
objects.  For example, \citet{censimcoe1997} use a large $10h^{-1}$
comoving Mpc box to estimate the numbers and sizes of discrete
\lya ``clouds'' as a function of overdensity.  For
overdensities of $\rho/\bar{\rho}=30$, they distinguish $\sim 10$ 
clouds/Mpc$^{-3}$, which have a median size of $R\sim 26$ kpc.
Comparing these numbers to the $n\sigma$ product in Equation
\ref{eqn_nsig_norm}, we find that the observed value is overestimated by
a factor of $\sim 15$.  However, the true spatial resolution of this
simulation at $z=2.5$ is only $\sim 30$ physical kpc, so the structures of 
interest are not well-sampled.  It is therefore possible that either the
average cloud cross section is overestimated, or structures that
would collapse to smaller highly overdense regions are smoothed by the
simulation resolution, resulting in an overestimate of the number
density.  

The alternative approach is taken by \citet{rauch1997}, who simulate a
much smaller $1.4h^{-1}$ comoving Mpc box at higher resolution to
study the assembly of proto-galactic clumps (PGCs) and small galaxy groups.
\citet{mulchaey1996} have also outlined the expected observational 
signature of these type of structures.  Their main
prediction is that quasar sightlines passing through small groups
should penetrate multiphase absorbers with strong \hi.  One of the phases
is hot, highly ionized, and seen principally in \ovi - this represents the
intra-group medium.  A second, cooler phase with lower ionization 
species (e.g. \civ, \siiv) is associated with the extended haloes of 
the galaxies themselves.  The agreement between these predictions and
our observations, as well as observations of galaxy clustering at
high redshift \citep{steidel1998,venemans2002}, lend merit to this
hypothesis. 

In \citet{rauch1997}, the authors note the size of PGC structures at a given
temperature, finding that gas with $T\ge 10^5$ K is found in 
spheroidal envelopes within $\sim 30$ kpc of PGCs at $z=3$.  
From their Figure 2, we estimate that $\sim 2-3$ such regions exist in 
a volume of approximately $0.23$ physical
Mpc (scaled to $z=2.5$).  From Equation \ref{eqn_nsig_norm} we see
that the implied $n\sigma$ product is again overestimated by a
factor of $\sim 15-20$.  However, the volume for this high resolution grid 
is not entirely representative, in that it was explicity chosen to
include several proto-galactic clumps and probably overestimates the
true global number density.  Even with this selection, the
small number of PGCs results in significant poisson error in $n$.

If the mapping between \ovi absorbers and structures in the
simulations is correct, and if the size and density estimates of the
simulations are accurate, the overestimate of the $n\sigma$ product
would argue against the widespread intergalactic distribution of Oxygen, 
as has been suggested in connection with Population III enrichment scenaria.  
However, the good agreement between simulations and observations at
low redshift suggests that inflow shocks are producing \ovi at some
level.  Lacking direct access to simulations, our comparisons here are
necessarily crude; a more sophisticated comparison with current and 
future generations of simulations is warranted to determine if the
discrepancies at high redshift are essentially physical or numerical.

\subsection{\ovi: Tracer of Galactic Outflows at High Redshift?}\label{sec_outflow} 

The second, equally plausible mechanism we consider for \ovi
production is the collisional ionization of hot winds expelled from
galactic environments.  A desirable feature of this model
is that it provides a natural explanation for the high
metallicities implied by the arguments from Section
\ref{sec_metallicity}.  Also, unlike the infall hypothesis whose motivation
is essentially theoretical, the outflow hypothesis draws upon
direct observations of galactic winds at both high and low 
redshift \citep{pettini2001, franx1997,dawson2002}.  

The existence of large-scale galactic outflows is a well-documented
phenomenon in the nearby universe, and is usually assocated with regions of
intense star-forming activity (see Heckman 2001 for a review).  Recent 
observations with the FUSE satellite have directly 
confirmed the presence of substantial \ovi in a supernova-driven
superbubble flowing out of the dwarf starburst NGC 1705
\citep{heckman2001}.  Many of this system's properties strongly resemble
those of the systems found in our survey.  These include the
wind velocity ($\Delta v\sim 100$ km/s), the gas metallicity
($[O/H]\sim -1.5$), and column density ($N_{\movi}\sim 10^{14.3}$).  
The physical gas density in the NCG 1705 wind is somewhat
higher than the range we have considered at $n_H\sim 0.03$, but since 
the \ovi in a galactic wind is dominated by collisional ionization, the 
gas density of the high redshift systems is essentially unconstrained
and could take on any value above $n_H\sim 2\times 10^{-5}$.  
The superbubble in NGC1705 appears to be in a
``breakout'' phase, where the shell of the bubble fragments in the
wake of a superwind.  The \ovi is produced in a thin layer near the 
interface of the fragmenting shell and the onrushing wind, which
is thought to be capable of escaping to large distances. 

There is some evidence that winds may be even more frequent at high 
redshift, as the outflow signature is a ubiquitous feature in the
spectra of Lyman Break galaxies at $z=3$ \citep{pettini2001}.  These winds 
are seen in interstellar absorption lines 
which are blueshifted by several hundred km/s from the \hii lines thought 
to trace the stellar population.  As a simple consistency test, 
if we compare the 
comoving number density  
of LBGs ($\phi_*=0.004$ Mpc$^{-3}$; Adelberger \& Steidel 2000)
with our estimates of the $n\sigma$ product for \ovi absorbers, we
find that at $z=2.5$ LBGs can account for all of the observed
\ovi absorption in our survey if they blow winds out to radii of 
$41$ kpc.  This is in close agreement with size estimates for the \ovi 
absorbers based upon pathlength considerations, and is easily achieved
in superwinds like those seen in low redshift star forming galaxies
\citep{heckman2001b}.  For winds moving at $100-200$ km/s, the
timescale required for a wind to entrain this volume is 
$\tau_{\rm wind}\sim 250-500$ Myr, or about $15\%$ of the Hubble 
time at $z=2.5$.  
This wind timescale is in agreement with the median 
star formation age of $320$ Myr found from
spectral fitting to the rest-frame optical colors of LBGs
\citep{shapley2001}.  Clearly, this result does not establish a
direct connection between LBGs and \ovi production, but it
demonstrates that known galaxy populations at high redshift could 
produce winds similar to those required to explain the properties of
\ovi.  

\section{Summary and Conclusions}\label{sec_conclusions}

We have performed a survey for \ovi absorption in the intergalactic 
medium at $2.2<z<2.8$ along the lines of sight to five quasars.
Eighteen \ovi systems were identified, 12 of which constitute the
primary sample for this paper.  
The remaining six systems are located near their respective background 
quasars, and are either ejected from the 
central engine or affected by the strong local radiation field.
At high redshift, blending from the \lya forest is a significant
limitation in \ovi searches.  Accordingly, we have provided
quantitative estimates of both the completeness of the sample and its 
contamination level.  The completeness - a measure of the fraction of
true \ovi systems that are detected (i.e. not blocked by the forest) - is found
to be $f_c\approx 41\%$.  The contamination level, which measures the
number of \lya line pairs in the forest masquerading as \ovi
doublets, is estimated at $\le 10\%$ of our identified \ovi systems.   

The intergalactic systems we have detected share the following
directly observable properties:
\begin{enumerate}
\item{They are located in highly overdense environments.  
These regions are characterized by strong \lya
absorption, either as a Lyman-limit system or as a collection of
several clouds with $N_{\mhi}\sim 10^{15.5}$.  One \ovi system is 
associated with a weak damped \lya absorber.}
\item{\civ absorption is
seen in the environment of every \ovi system, and other lower
ionization species (\siiv, \cii, \siiii, \siii) are often present as
well.}
\item{The detailed velocity structure of \ovi does not generally match
that of the other heavy elements.  The \ovi profiles typically show
less substructure than \civ or \siiv, and are sometimes offset from
lower ionization lines by several hundred km/s.  
}
\end{enumerate}

Surprisingly, we have failed to detect any \ovi lines associated with 
lower density regions of the \lya forest.  Cosmological simulations
\citep{hellsten1998,rauch1997} predict that photoionized \ovi should be 
an efficient tracer of metals in clouds with $13.5 <
\log(N_{\mhi}) < 14.5$, yet we detect no intergalactic \ovi lines near 
regions with \hi column densities below $\log(N_{\mhi})=14.6$.  
The survey data were originally taken to the standards 
required to detect \ovi in forest lines with [O/H]=-2.5, and we
find several clean systems where \ovi should be seen in the
data at this level but is not.  However, there are many systems where
we would not detect weak \ovi due to forest blending or noise, and 
it remains to be seen whether this finding is statistically 
significant or is simply a reflection of cosmological scatter in the 
Oxygen abundance or ionizing radiation field.
These questions will be addressed in forthcoming work.  

To aid in our interpretation of the detected systems, we have
performed a suite of ionization simulations using the CLOUDY software
package.  Runs were made for the photoionization equilibrium case, and
also at a range of fixed temperatures between $10^4-10^6$ K to investigate
the onset of collisional \ovi production.  From comparisons between
the observations and ionization calculations, we have assembled a
basic physical description of high redshift \ovi systems that
may be summarized as follows:
\begin{enumerate}
\item{Their physical extent and gas density may be conservatively 
constrained at $L\le 200$ kpc and $\rho/\bar{\rho}\ge 2.5$.  This
was calculated by comparing the maximum observed \ovi
linewidth with the broadening expected for clouds of different sizes
due to the Hubble flow.  For the median observed value of 
$b_{\movi}=16$ km/s, the inferred cloud sizes 
and densities are $L\sim 60$ kpc and $\rho/\bar{\rho}\sim 10-30$.}
\item{They possess at least two distinct gas phases.  One of these
gives rise to absorption in photoionized \civ and \siiv, and has
temperatures in the range $T=20,000-40,000$ K, and overdensities of
$\rho / \bar{\rho}\sim 500$.  This is the same variety of gas which
gives rise to the common \civ and \siiv absorption seen in all high
redshift quasar spectra.  The second phase is physically distinct,
and traced only in \ovi absorption.  Its temperature is difficult to 
constrain because of uncertainties in the nonthermal contribution to line
broadening.  However, the distribution of $T_{\rm max}$ shown in Figure
\ref{fig_b_temp} indicates that the \ovi temperature structure differs
from that of \civ and \siiv, and favors higher temperatures where
collisional production of \ovi would be significant.}
\item{They are strongly clustered on velocity scales of $\Delta v=100-300$
km/s, and show weaker clustering signal out to $\Delta v = 750$ km/s.
The power law slope of the two-point correlation function is similar to
that seen from local galaxy and cluster surveys, with a comoving
correlation length of $\sim 11h_{65}^{-1}$ Mpc, intermediate between galactic 
and cluster scales.  While the correlation at large velocities could 
be the signature of galactic winds, a simple geometric wind model does not
accurately predict the shape of the TPCF over the entire velocity
range where signal is seen.  We have argued that the signal in the \ovi TPCF
is dominated by the signature of large scale clustering.  This does not
rule out the possibility that its shape is slightly modified by
peculiar motions or outflows.}
\item{On average, they possess Oxygen abundances of $[O/H]\ge
-1.5$, about 10 times higher than the level observed in the general
IGM.  This level of enrichment probably requires additional metal
input above the level supposed to originate from Population III
stars.  This conclusion assumes that the \ovi absorbers contain $\le
10\%$ of the baryons at high redshift, i.e., that they are distinct
from the more tenuous IGM that gives rise to the \lya forest and
contains $\ge 90\%$ of the baryons.}
\item{The integrated mass of warm-hot gas at $z\sim2.5$ amounts
to $\ge 0.5\%$ of $\Omega_b$.  This may be compared with the latest
estimates of \citet{savage2002}, who measure $\Omega_{\rm WH}$ at low
redshift to be only $25\%$ higher using identical assumptions.  
If the average metallicity of the universe increases with time, the
difference is even less pronounced.  This relatively weak evolution 
may indicate that \ovi is seen only during a short phase in the 
cooling history of pre-heated gas.  In this case, the balance of the
shock heated IGM, integrated over time, could reside 
in gas with $T>10^6$ K, which has long cooling timescales and 
is not efficiently traced by \ovi.}
\end{enumerate}

Based on the absorption data alone, we cannot make a strong
distinction between the scenario where \ovi is produced in
wind-induced shocks, and the scenario where it is produced in
accretion shocks due to structure formation.  The high average
metallicity we have measured in the \ovi absorbers seems to point 
towards the wind hypothesis, though we cannot measure metallicities 
on a system-by-system basis.  
The overdensities we associate with the observed \ovi
systems are between those observed in the \lya forest and collapsed
structures.  At the low end of the allowed density range, \ovi with
properties similar to those observed can be produced through either 
collisional ionization or photoionization, though some systems can
only be explained by collisional processes even at low density.  At
higher densities than $n_H\sim 3\times 10^{-4}$ collisional
ionization is required to explain the strength of \ovi relative to
other ions for solar relative abundances.

We have compared the \ovi absorbers to structures of similar 
overdensity and temperature in cosmological simulations to test 
the plausibility of the accretion hypothesis.  Using estimates of the
number density and cross section of these structures, we find that 
the number of \ovi detections predicted for
the survey is too high by a factor of $\sim 15$.  However, the 
comparison is not always straightforward because
simulations in the literature typically contain either sufficient spatial 
dynamic range to resolve the scales of interest, or a large enough 
volume to minimize cosmic variance - but not both.  The generation of 
simulations currently running should be able to address
this question more accurately, to distinguish whether this
discrepancy is physical or numerical.

A critical assessment of the wind model is also
difficult, as current cosmological simulations are not capable of
treating such a complex processes in full physical detail.  
However, we have compared our results with the most well studied 
population of wind-producing galaxies at high redshift, the Lyman
break galaxies.  Using current estimates of the LBG comoving number 
density, we find that they are capable of producing all of the observed \ovi 
absorption if each LBG drives winds to a radius of $\sim 41$ kpc -
similar to the size inferred from our pathlength analysis.  
Furthermore, the time required to drive winds to this distance at
$v_{\rm wind}=100-200$ km/s is in agreement with recent estimates of 
the star formation ages of the LBGs.  While this coincidence does 
not constitute
a direct connection between LBGs and \ovi absorbers, it does
demonstrate that known galaxy populations could plausibly give rise to
the amount of \ovi seen in our survey.

Ultimately, much of the warm-hot gas that has
been suggested as a baryon reservoir at low redshift may be
hotter than $10^6$ K, and hence be undetectable in \ovi.  
The fact that $T_{\rm max}$ for \ovi spans the entire 
range from $10^5 < T < 10^6$ K without any clear peak suggests that
the high temperature limit for \ovi lines results from further
ionization of the Oxygen, rather than the actual detection of a 
maximum gas temperature.  Gas which is shocked to temperatures of 
$2\times 10^6$ K (near the peak of the predicted warm-hot temperature
distribution from Dav\'e et al 2001) 
at $\rho/\bar{\rho}\sim 5$ would have a cooling
timescale of $\tau_{\rm cool}\sim 1.5$ Gyr, which amounts to $60\%$ of 
the Hubble time at $z=2.5$.  At the same density, gas in the \ovi 
temperature range cools $\sim 30$ times faster, owing to its lower 
initial energy, and its location at the peak of the cooling curve.  
If this is the case, then
\ovi absorption would trace the ``tip of the iceberg'' with
respect to the total amount of warm-hot gas produced over cosmic
time.  In principle, the prospect of detecting the rest of the gas is
better in \ovii or \oviii X-ray absorption, but the number of
extragalactic objects bright enough for high resolution X-ray
absorption spectroscopy on current instruments is small, and all are
at low redshift.  The indications from the \ovi data, along with a 
handful of \ovii and \oviii detections, suggest that this hot gas exists, 
but it may be some time before it can be observationally 
characterized in a statistically robust sense.  

Since the lifetime of gas in the \ovi state is short, sources
of energy input are needed to produce and maintain the high state of
ionization in these systems.  Hence, the primary utility of \ovi may 
not be for measuring the total content of the warm-hot intergalactic 
medium, or tracing the metal content of the lowest density regions of
the forest, but rather for probing physics at the interface between 
galaxies and the IGM.  This connection has been suggested by the association
of galaxy groups and \ovi absorption at low redshift \citep{savage2002}; an
analogous connection at high redshift could aid in the
characterization of processes thought to have significant impact on
the thermal and chemical history of the IGM.

\acknowledgements
We thank Bob Carswell for assistance with VPFIT, Gary Ferland for
making CLOUDY available to the community, and Tom Barlow for
assistance with MAKEE.  We further acknowledge useful discussions with 
R. Carswell and J. Schaye regarding
the differences in our \ovi samples.  We also thank the Keck
Observatory staff for their assistance with the observations.  Finally,
we extend special thanks to those of Hawaiian ancestry on whose
sacred mountain we are privileged to be guests.  Without their
generous hospitality, the observations presented herein would not have
been possible.  W.W.S. and R.A.S. have been supported by NSF grant
AST-9900733.  R.A.S. further acknowledges partial support from the 
Lewis A. Kingsley Foundation.  M.R. is grateful for support from the
NSF through grant AST-0098492 and from NASA through grant AR-90213.01-A.

\appendix

\section{Comparison with Other Recent \ovi Surveys}
During the revision stage of this manuscript, Carswell, Schaye \& Kim
(2002, hereafter CSK) have reported on a similar search for \ovi 
in the spectra of $z\sim 2.5$
quasars, with the seemingly contradictory finding that the majority of
\ovi absorbers are low density, photoionized systems in the
\lya forest.  However, close examination of the two datasets
reveals that the differences in interpretation are due largely to the
selection criteria used to define our respective \ovi 
samples.  Given the large amount of \lya forest absorption at these
redshifts, there are two different but useful approaches to \ovi
sample definition.  We have chosen selection criteria designed 
to minimize the
number of false positive identifications in our final sample at the
expense of completeness, whereas CSK have chosen to minimize the 
incompleteness of their sample at the expense of a higher 
contamination level.  

From comparison of the two samples, it seems
that there may be two varieties of high redhift \ovi absorbers.  The
first, traced by our selection criteria, shows strong \hi absorption
and multiphase structure, and is probably formed near the
interface between high-redshift galaxies and the IGM.  The strongest 
1 or 2 systems in CSK's sample are of this nature.  
This is roughly the number of strong systems one would expect
to find in their survey from a simple rescaling of our results to the 
pathlength of their sightlines.  In addition,
CSK find many more weak \ovi systems that did not meet our
selection criteria, principally because one or both components of the
doublet were blended with a \lya or higher order \hi line.  The
properties of these systems are consistent with an origin in the 
tenuous regions of the \lya
forest.  We are currently engaged in a separate search for weak
\ovi absorption in our dataset; we therefore defer discussion on 
this topic to future work.  Here, we simply note that the two studies 
may actually complement each other by
sampling different segments of the total population of \ovi absorbers.

\clearpage
\begin{deluxetable}{c c c c c }
\tablewidth{0pc}
\tablecaption{Table of Observations}
\tablehead{{Object} & {Total $\lambda$ coverage (\AA)} & {$z_{em}$} & {$\Delta z_{\small OVI}$\tablenotemark{1}} & {$\Delta X_{OVI}$\tablenotemark{1,2}}}

\startdata

Q1009+2956 &	3200-6075 & 2.62 & 2.10-2.56  & 1.48 \\
Q1442+2931 &	3200-6150 & 2.63 & 2.19-2.57  & 1.24 \\
Q1549+1919 &	3160-6084 & 2.83 & 2.06-2.77  & 2.31 \\
Q1626+6433 &	3300-6180 & 2.32 & 2.15-2.26  & 0.36 \\
Q1700+6416 &	3250-6140 & 2.72 & 2.20-2.66  & 1.51 \\
\\
\hline
Total:	   &		  &	 & & 6.90 \\

\enddata
\tablenotetext{1}{\small Corrected to exclude regions within 5000 km/s of the
QSO emission redshift.}
\tablenotetext{2}{\small Does not include the effects of spectral
  blockage; see Section \ref{sec_xsection} for further discussion of
  this point.}

\end{deluxetable}
\begin{deluxetable}{c c c c c c c c c }
\tablewidth{0pc}
\tablecaption{Summary of Observed Systems}
\tablehead{{Sightline} & {$z_{abs}$} & {${\rm N}_{{\rm comp}, \movi}$} &
{$\log(N_{\movi,tot})$} & {${\rm N}_{\mhi>14.0}$} & {$\log(N_{\mhi,
max})$} & {$\Delta v_{\rm QSO}$} & {$\Delta v_{\movi-\mciv}$} & {Fig.}}

\startdata

Q1009+2956 & 2.253 & 4 & 14.626$\pm$ 0.014  & 2     & 17.806   &   $>10,000$ & 91.1 & 1  \\
	   & 2.429 & 2 & 13.602$\pm$ 0.030  & 3     & 17.687   &   $>10,000$ & 14.3 & 2 \\
	   & 2.606\tablenotemark{1} & 1 & 12.709$\pm$ 0.083  & 1  &   14.442  &  1,162     & \nodata \tablenotemark{4} & 3  \\
\\
Q1442+2931 & 2.439 & 11 & 15.001$\pm$ 0.014 & $3$  & $\sim 19.500$   & $>10,000$ & $\approx 0$ & 4 \\
	   & 2.623\tablenotemark{1} & 2  & 13.592$\pm$ 0.034 & 6  &  15.758   & $3,300$  & 14.2 & 5  \\
\\
Q1549+1919 & 2.320 & 8 & 14.506$\pm$ 0.140  & 3  & 15.195   & $>10,000$ & 9.6 &6\\
	   & 2.376 & 2 & 13.924$\pm$ 0.021  & 5  & 15.545   & $>10,000$ & 772 &7\\
	   & 2.560 & 1 & 13.564$\pm$ 0.096  & 2  & 15.219   & $>10,000$ & 336 &8\\
	   & 2.636 & 2 & 13.250$\pm$ 0.021  & 1  & 15.220   & $>10,000$ & 5.4 &9    \\\
	   & 2.711\tablenotemark{2} & \nodata & \nodata    & \nodata   & \nodata & 9,460 & $\sim 0$ &10\\
\\
Q1626+6433 & 2.245 & 2 & 14.827$\pm$ 0.032  & 2  & 15.502 & 7,317  & 1.3 &11\\
	   & 2.321\tablenotemark{1} & 1 & 14.256$\pm$ 0.024  & 4  & 15.423   & -300 &  37.8 &12 \\
\\
Q1700+6416 & 2.316 & $\ge 4$ & 14.968$\pm$ 0.026   & 2  & 17.623  & $>10,000$ & 13.7 &13\\
	   & 2.379 & 1 & 13.542$\pm$ 0.028  & 2  & 15.405  & $>10,000$ & 6.8 &14\\
	   & 2.436 & 9 & 14.426$\pm$ 0.081 & 7  & 16.943   & $>10,000$  & $\approx 10$ &15\\
	   & 2.568 & 2 & 13.587$\pm$ 0.039 & 1  & 14.552 & $>10,000$ & 7.5   &16\\
	   & 2.716\tablenotemark{2} & 4 & \nodata & \nodata  & \nodata   & 2250 & 3.0  &17 \\
	   & 2.744\tablenotemark{1} & 3 & 13.782$\pm$ 0.058   & 0  & 13.840   & 1500 & \nodata \tablenotemark{4} &18 \\

%
\enddata

\tablenotetext{1}{\small Excluded from cosmological statistics because
of proximity to background quasar.}
\tablenotetext{2}{\small Ejected system.}
\tablenotetext{3}{\small Column density and component structure for
\hi taken from \citet{burles1998}}
\tablenotetext{4}{\small No \civ detected}
\end{deluxetable}

\clearpage

\begin{figure}
\figurenum{1}
\epsscale{0.75}
\plotone{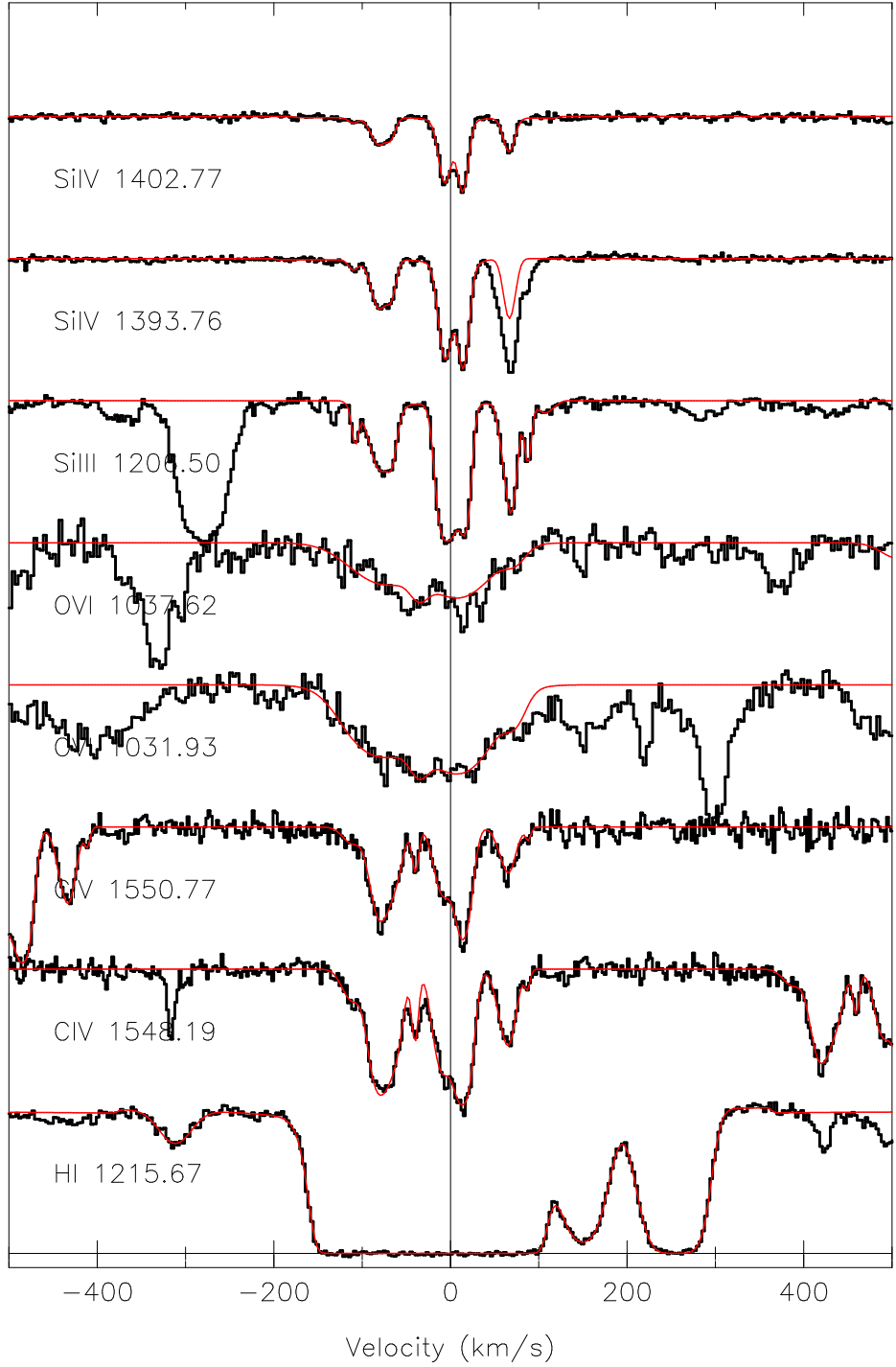}
\caption{ Stacked velocity plot of the $z=2.253$ system in Q1009+2956 
(Described in Section \ref{sec_1009_2.25} of the text).  The thin
solid line in this figure and all following velocity plots shows the
best-fit model obtained from the VPFIT profile fitting software.}
\end{figure}

\begin{figure}
\figurenum{2}
\epsscale{0.75}
\plotone{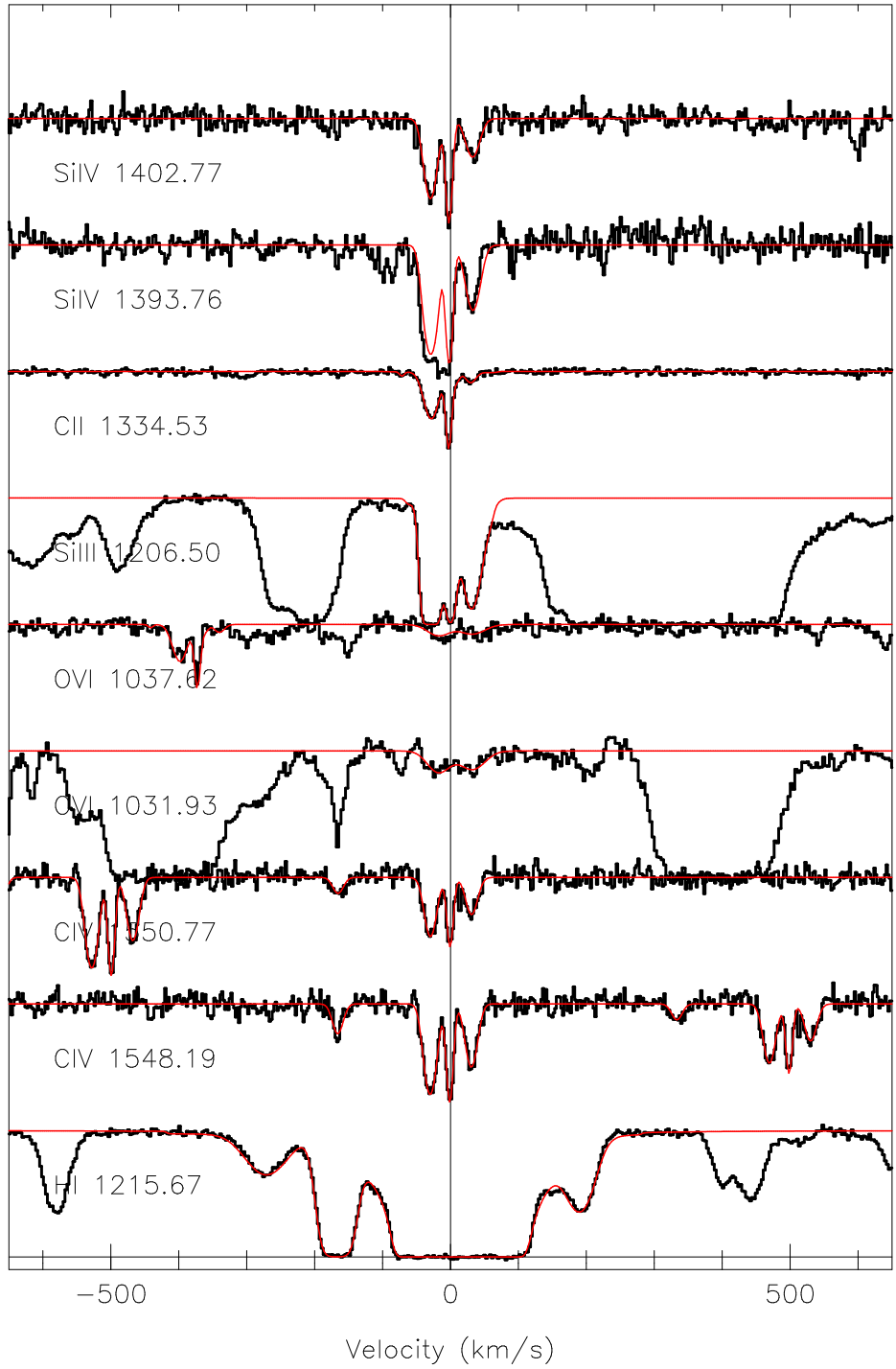}
\caption{ Stacked velocity plot of the $z=2.429$ system in Q1009+2956 
(Described in Section \ref{sec_1009_2.42} of the text).}
\end{figure}

\begin{figure}
\figurenum{3}
\epsscale{0.75}
\plotone{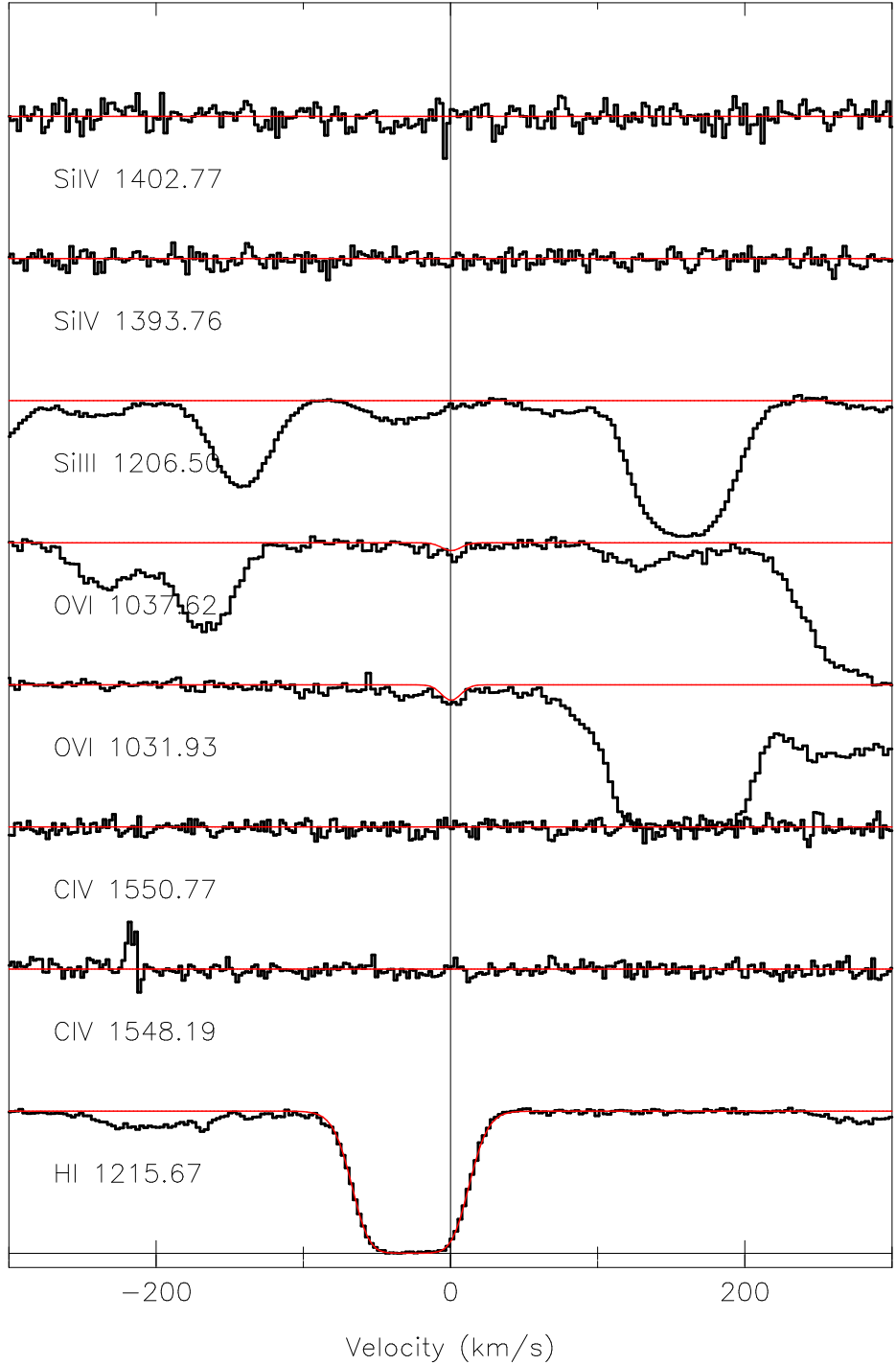}
\caption{ Stacked velocity plot of the $z=2.606$ system in Q1009+2956 
(Described in Section \ref{sec_1009_2.60} of the text).  This system
has been excluded from the cosmological statistics due to its
proximity to the background quasar.}
\end{figure}

\begin{figure}
\figurenum{4}
\epsscale{0.75}
\plotone{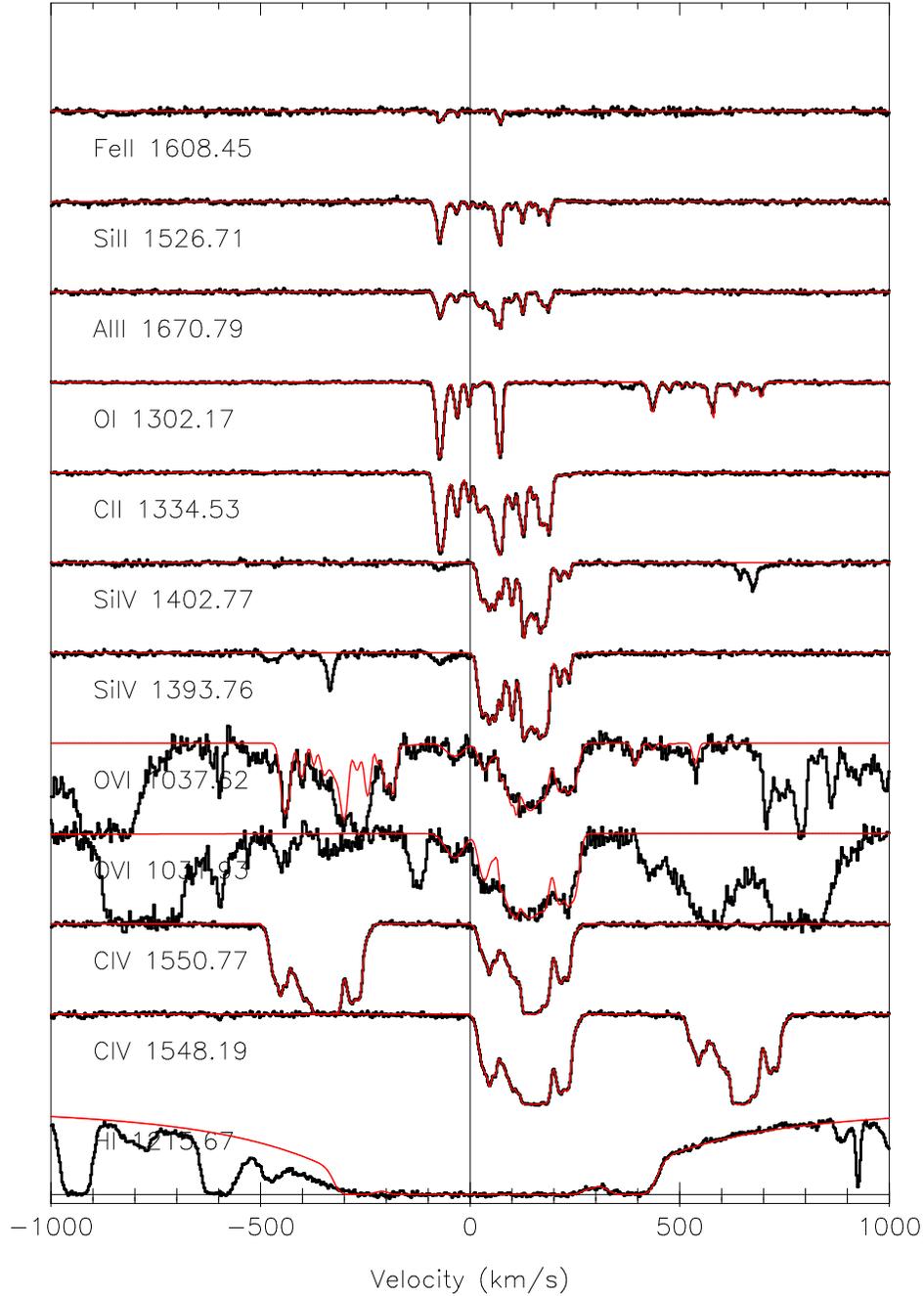}
\caption{ Stacked velocity plot of the $z=2.439$ system in 
Q1442+2931, which is a weak DLA and the strongest system in the survey 
(Described in Section \ref{sec_1442_dla} of the text).
The additional absorption shown in the \ovi 1037\ang fit is due to the
1036\ang transition of \cii.  Likewise, the absorption at +500 km/s in
the \oi profile is caused by \siii 1304\AA.}
\end{figure}

\begin{figure}
\figurenum{5}
\epsscale{0.75}
\plotone{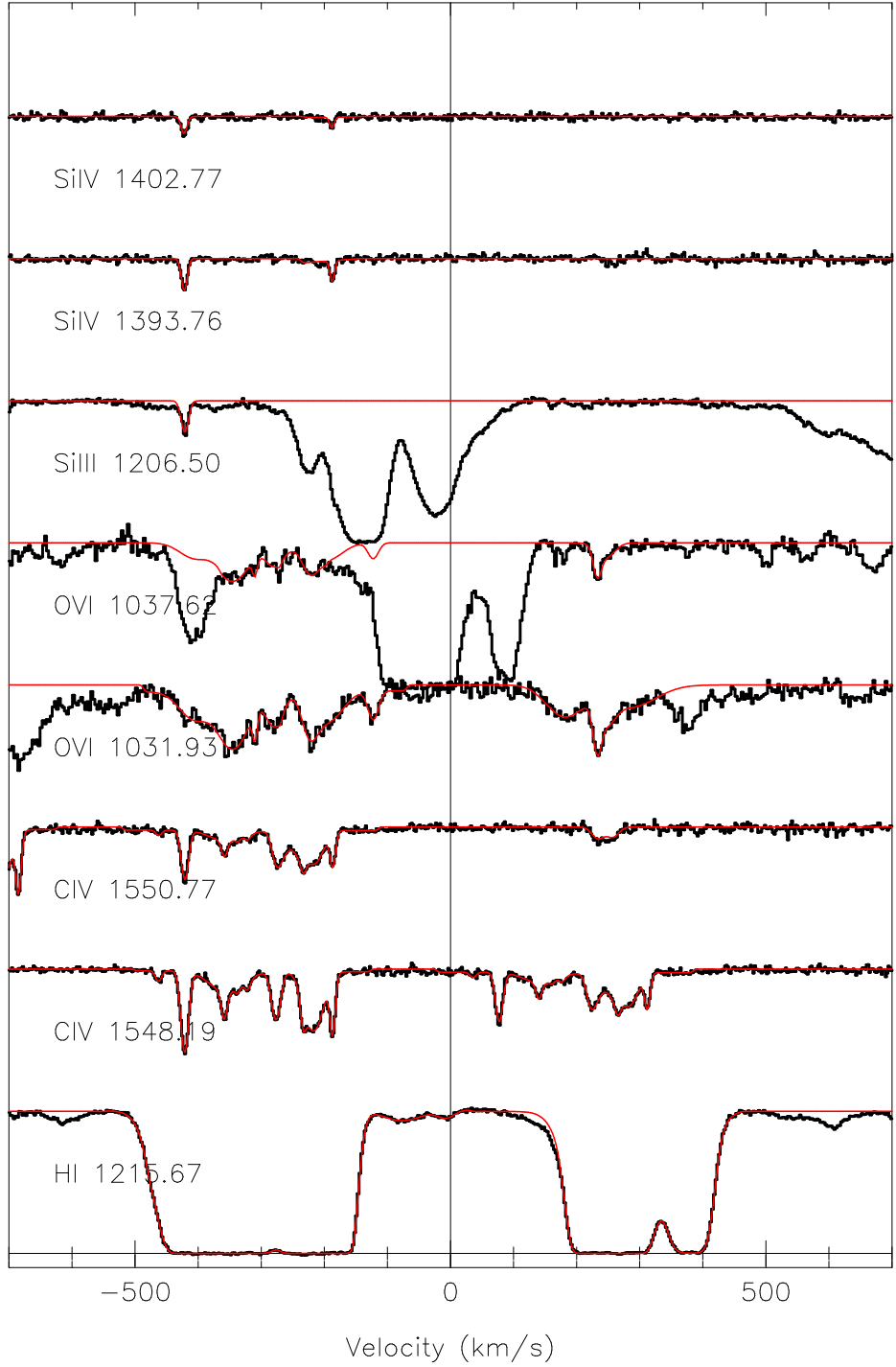}
\caption{ Stacked velocity plot of the $z=2.623$ system in 
Q1442+2931 (Described in Section \ref{sec_1442_2.62} of the text).  This system
has been excluded from the cosmological statistics due to its
proximity to the background quasar.}
\end{figure}

\begin{figure}
\figurenum{6}
\epsscale{0.75}
\plotone{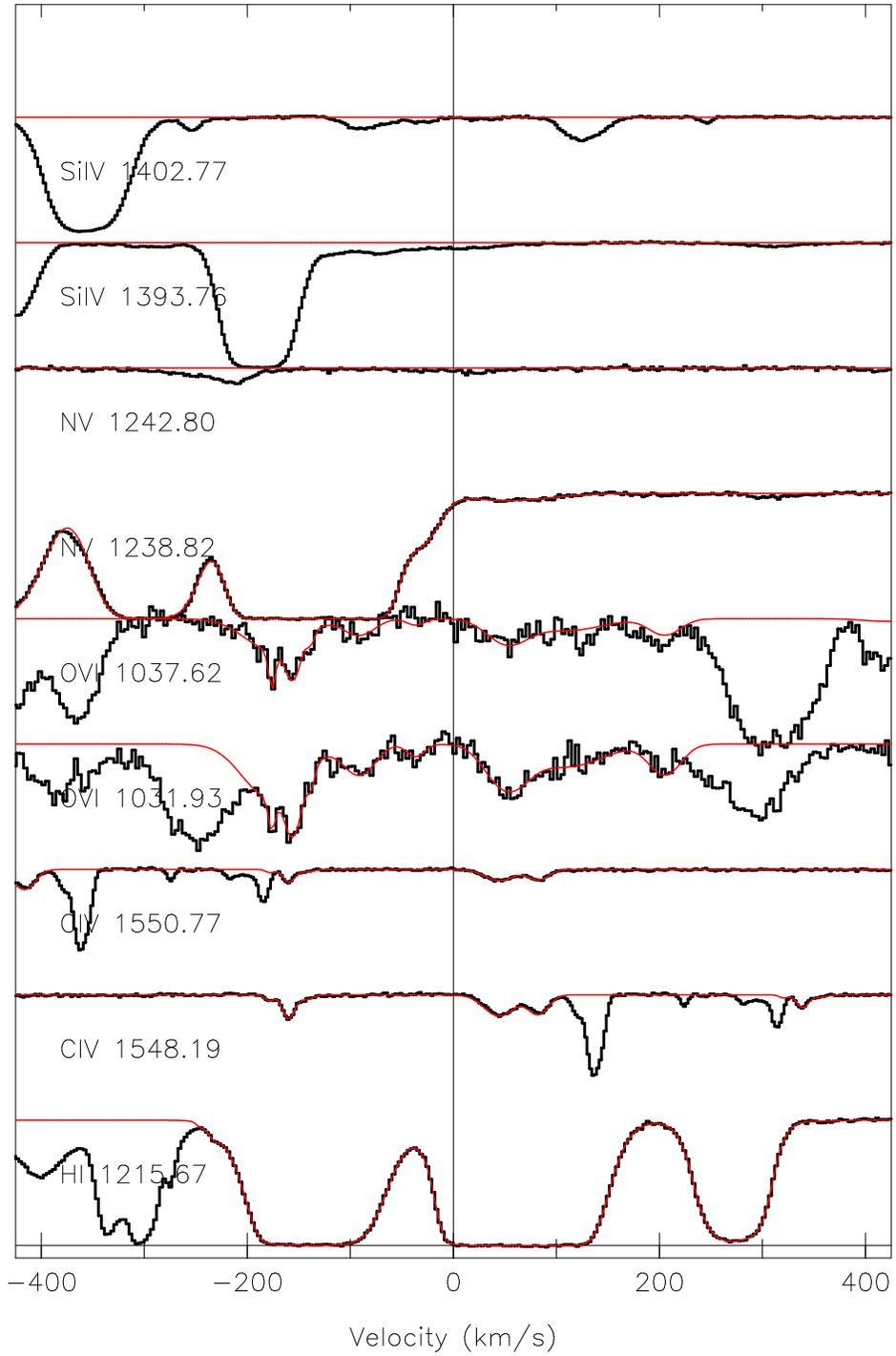}
\caption{ Stacked velocity plot of the $z=2.321$ system in Q1549+1919 
(Described in Section \ref{sec_1549_2.32} of the text).  The \ovi 1032\ang
profile has been adjusted to remove a \lyb line which was blended over
the range 0-200 km/s in the figure.}
\end{figure}

\begin{figure}
\figurenum{7}
\epsscale{0.75}
\plotone{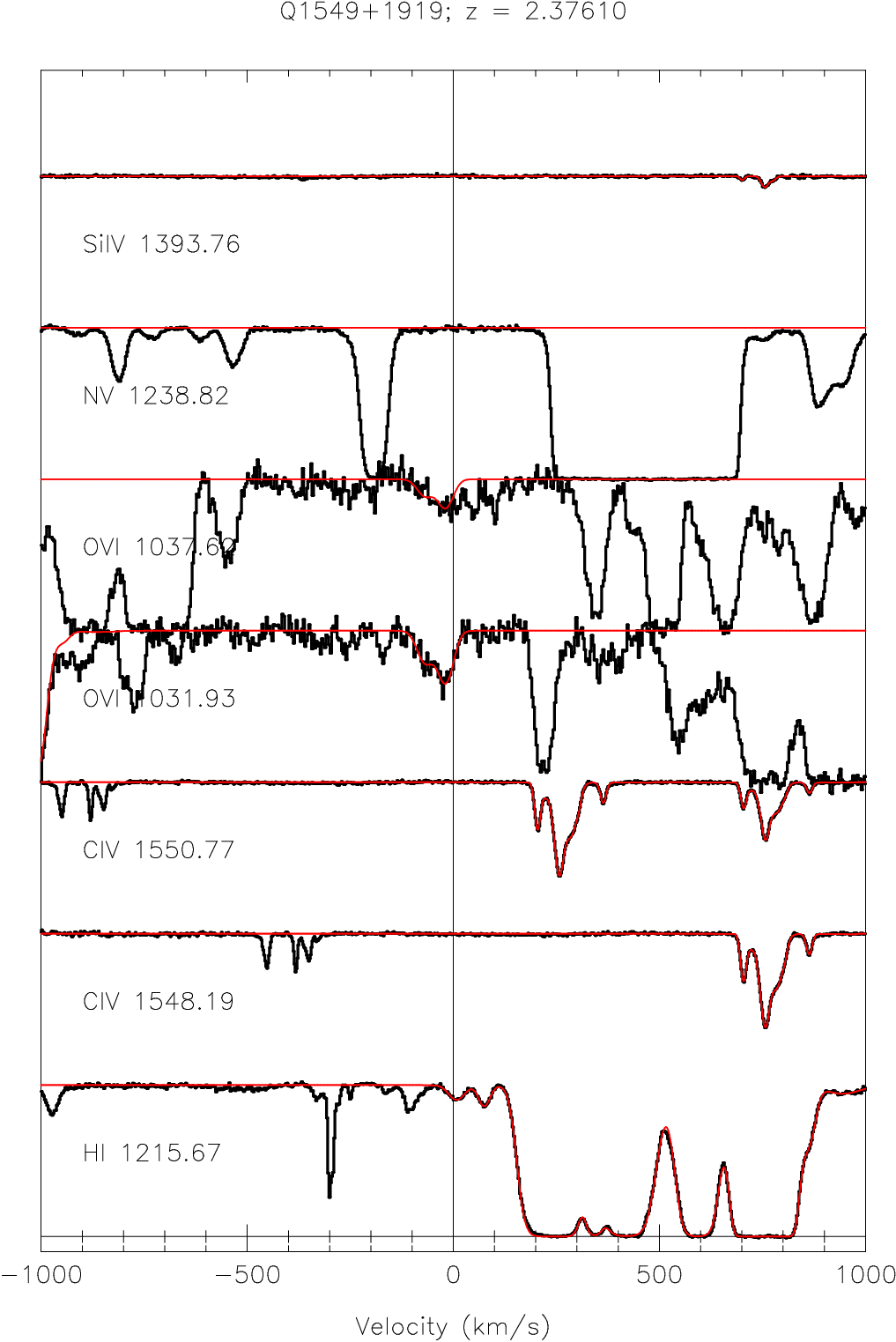}
\caption{ Stacked velocity plot of the $z=2.375$ system in Q1549+1919 
(Described in Section \ref{sec_1549_2.37} of the text).}
\end{figure}

\begin{figure}
\figurenum{8}
\plotone{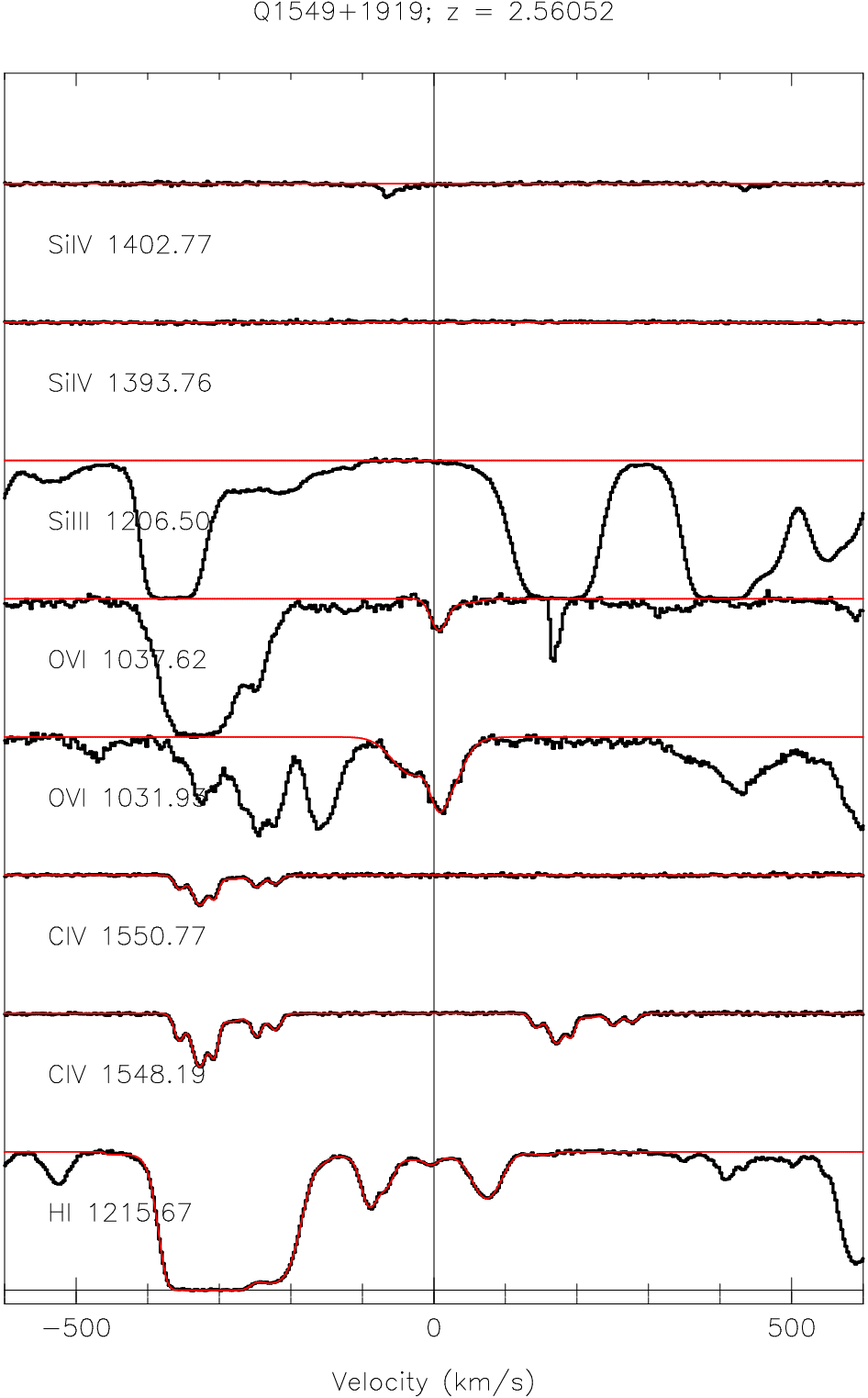}
\caption{ Stacked velocity plot of the $z=2.561$ system in Q1549+1919 
(Described in Section \ref{sec_1549_2.56} of the text).}
\end{figure}

\begin{figure}
\figurenum{9}
\plotone{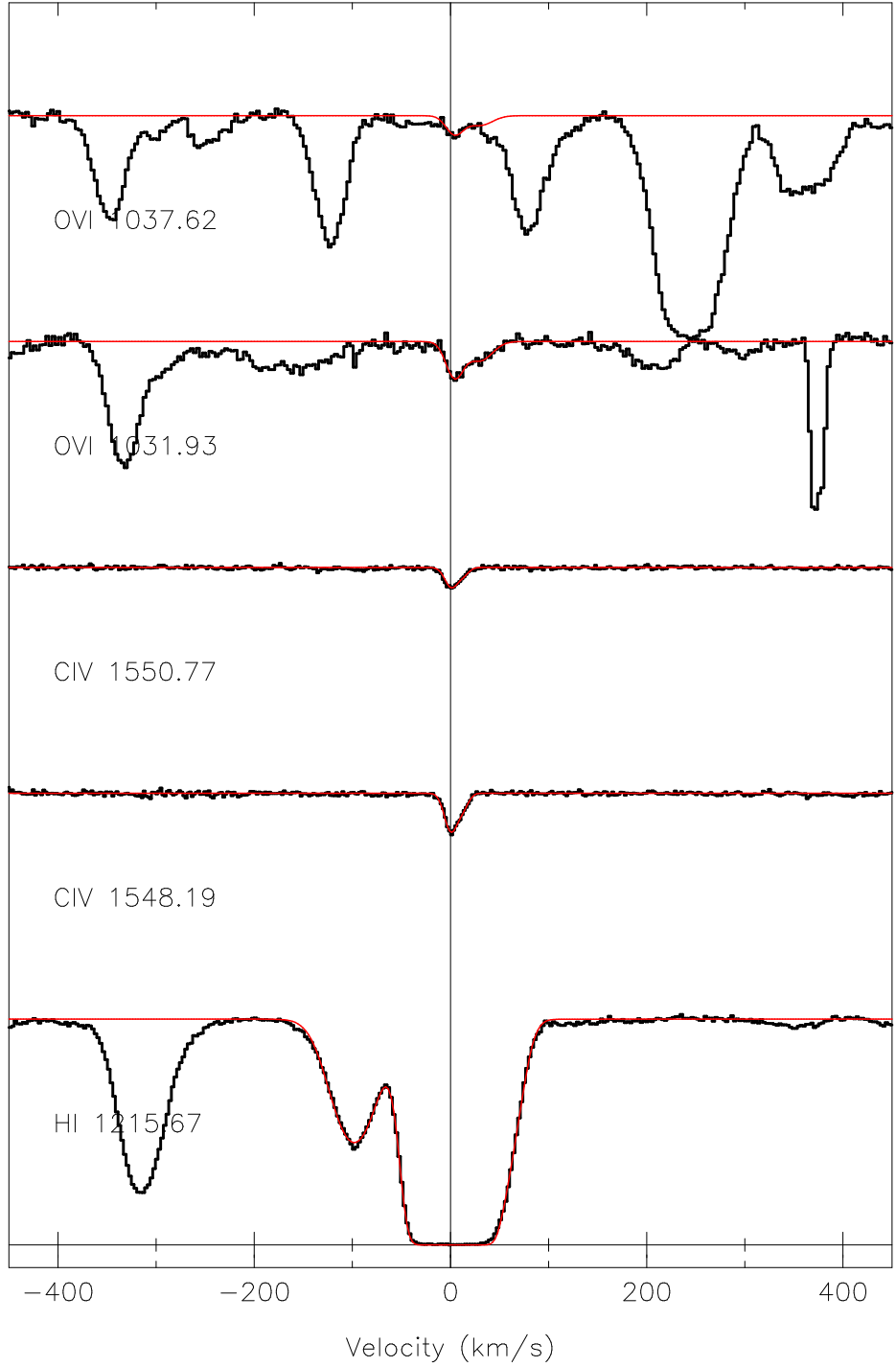}
\caption{ Stacked velocity plot of the $z=2.63$ system in Q1549+1919 
(Described in Section \ref{sec_1549_2.63} of the text).}
\end{figure}

\begin{figure}
\figurenum{10}
\epsscale{0.75}
\plotone{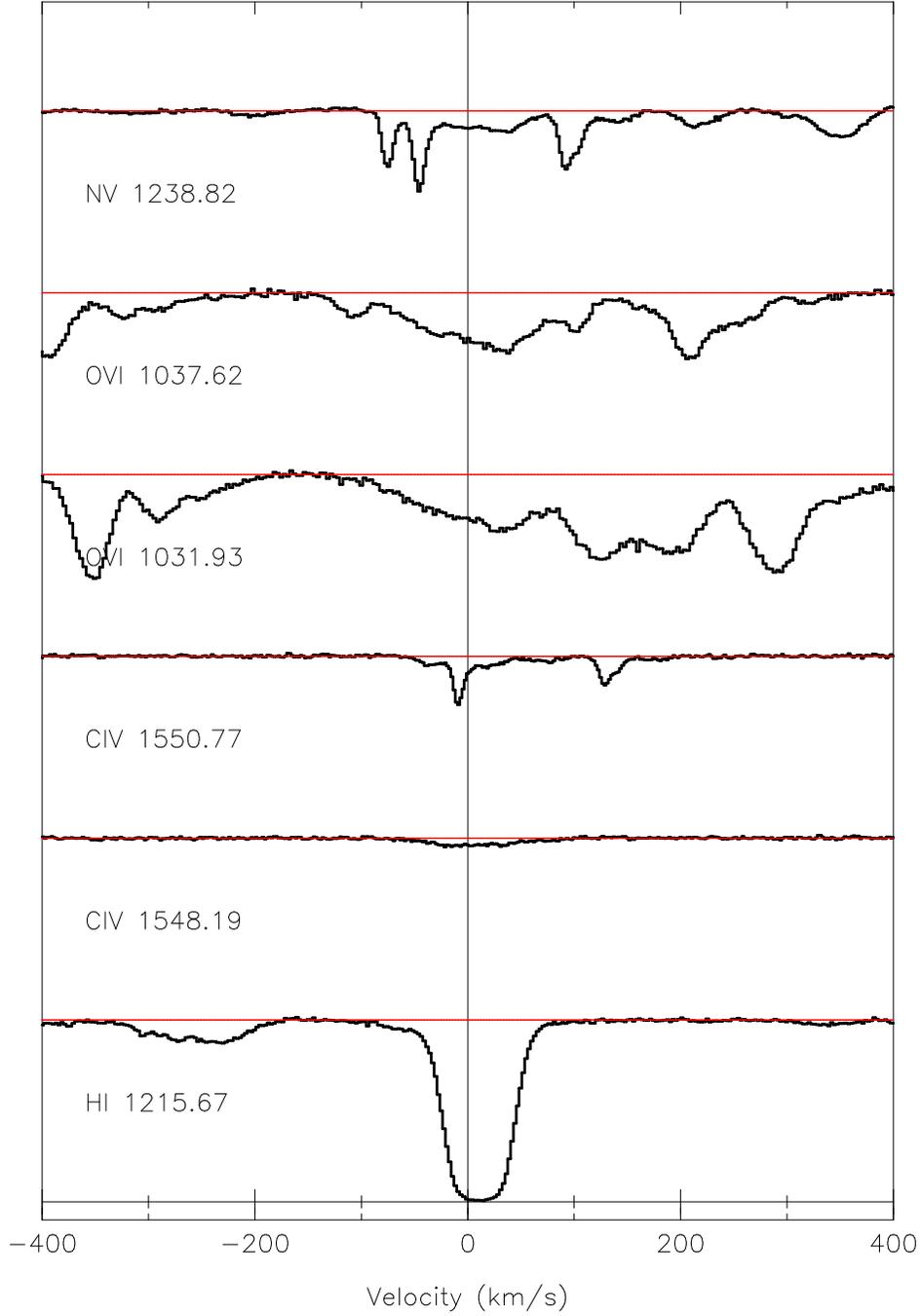}
\caption{ Stacked velocity plot of the $z=2.711$ system in Q1549+1919 
(Described in Section \ref{sec_1549_2.71} of the text).  Because the
\ovi doublet ratio is unity over the whole profile, we classify this
as an ejected system which is partially covering the continuum source.}
\end{figure}
\clearpage

\begin{figure}
\figurenum{11}
\epsscale{0.75}
\plotone{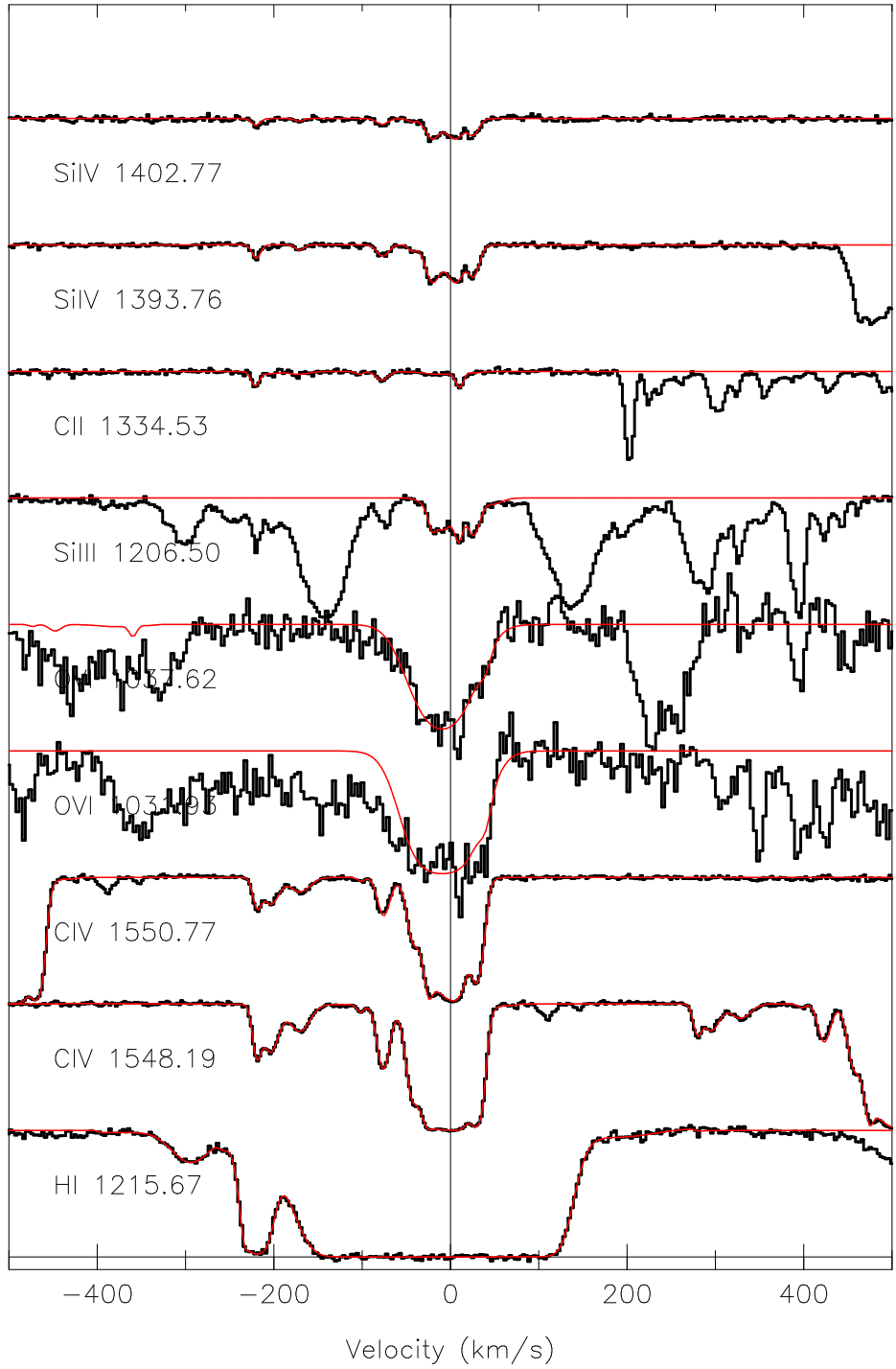}
\caption{ Stacked velocity plot of the $z=2.245$ system in Q1626+6433 
(Described in Section \ref{sec_1626_2.24} of the text).}
\end{figure}

\begin{figure}
\figurenum{12}
\epsscale{0.75}
\plotone{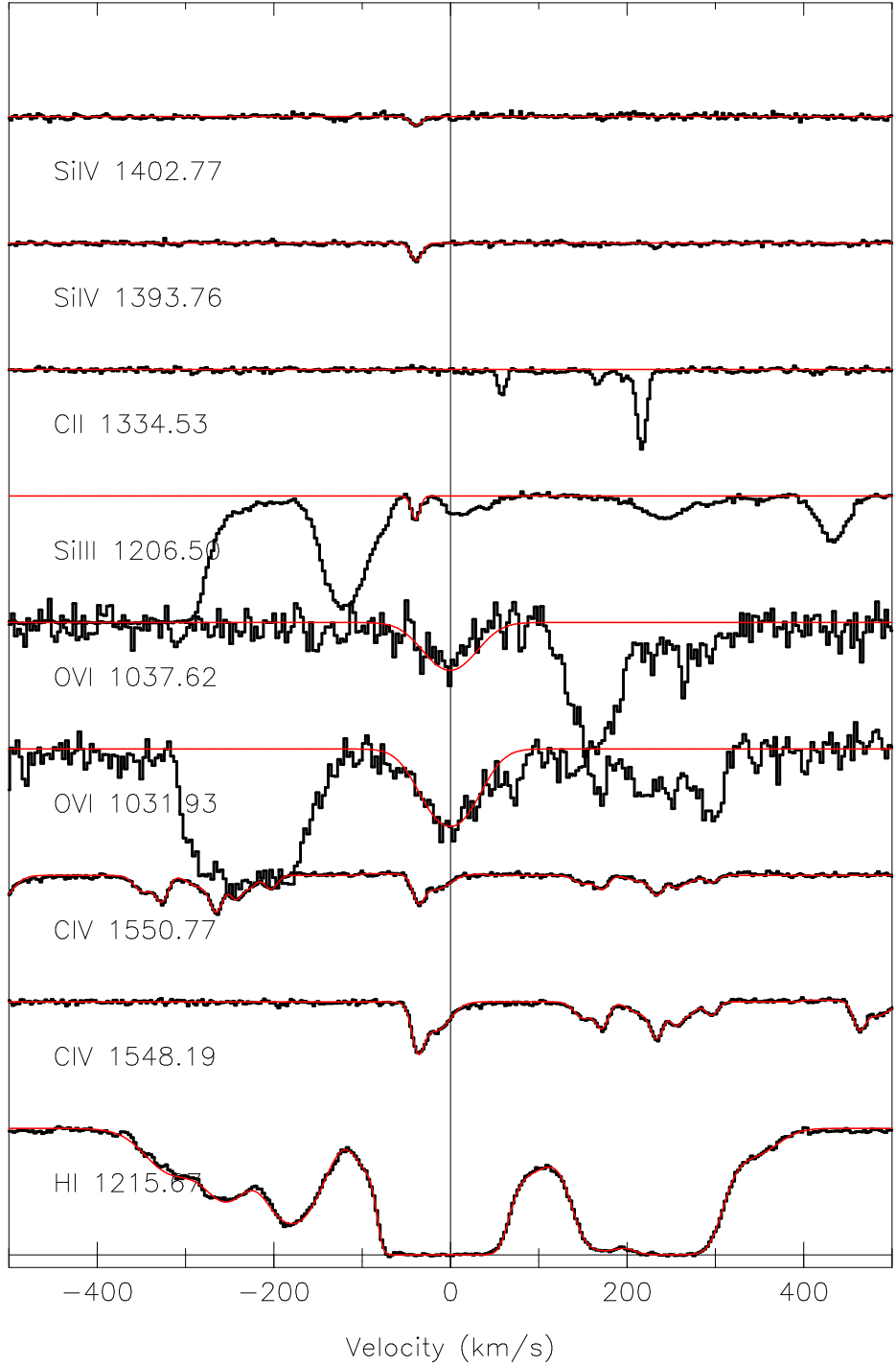}
\caption{ Stacked velocity plot of the $z=2.321$ system in Q1626+6433 
(Described in Section \ref{sec_1626_2.32} of the text).  This system
is located 200 km/s {\em redward} of the emission redshift of the QSO
(measured from \lya).  It is therefore excluded from the cosmological
statistics, though we believe it is likely caused by the QSO host
galaxy itself or a very nearby companion.}
\end{figure}

\begin{figure}
\figurenum{13}
\epsscale{0.75}
\plotone{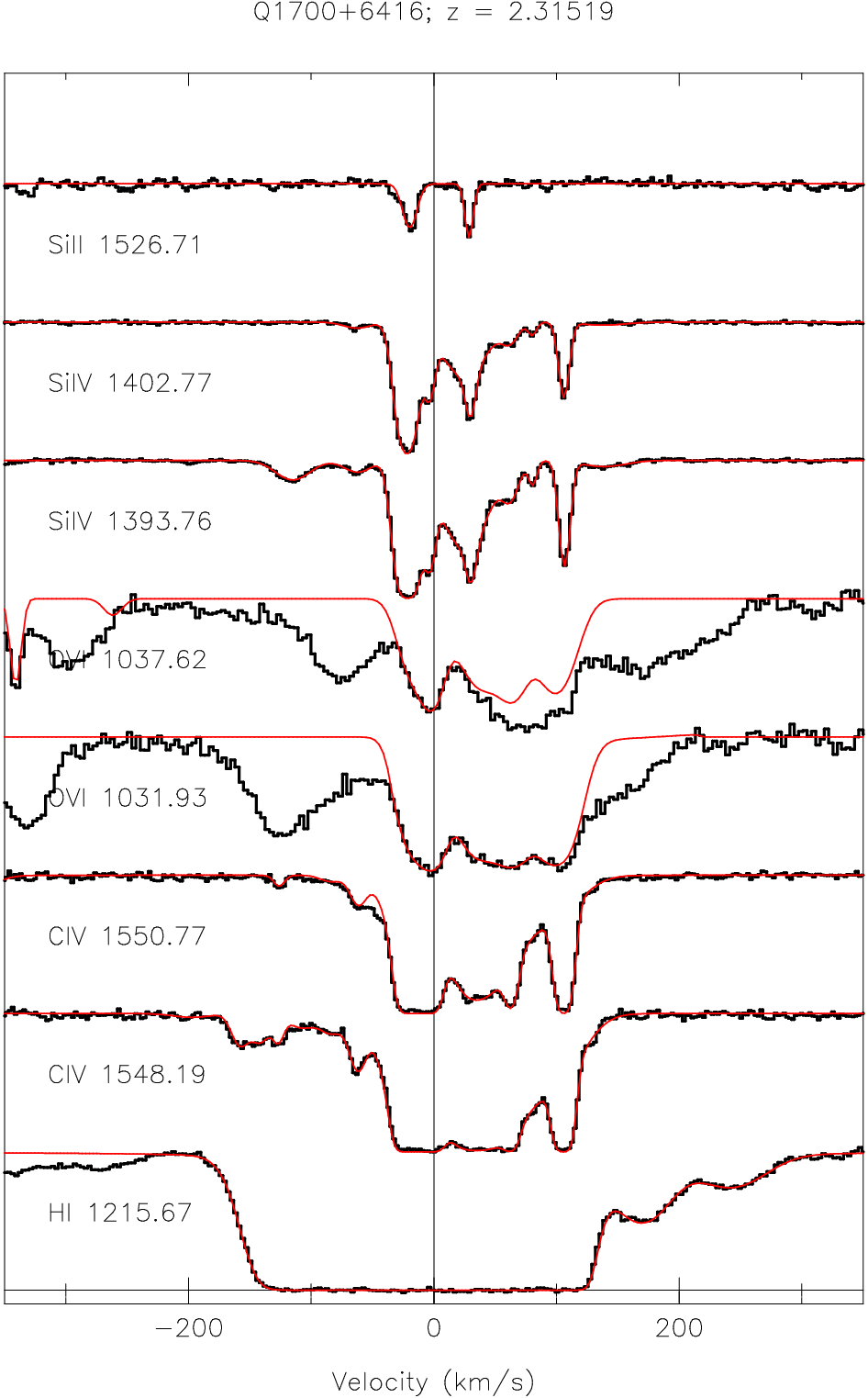}
\caption{ Stacked velocity plot of the $z=2.315$ system in Q1700+6416 
(Described in Section \ref{sec_1700_2.31} of the text).}
\end{figure}

\begin{figure}
\figurenum{14}
\epsscale{0.75}
\plotone{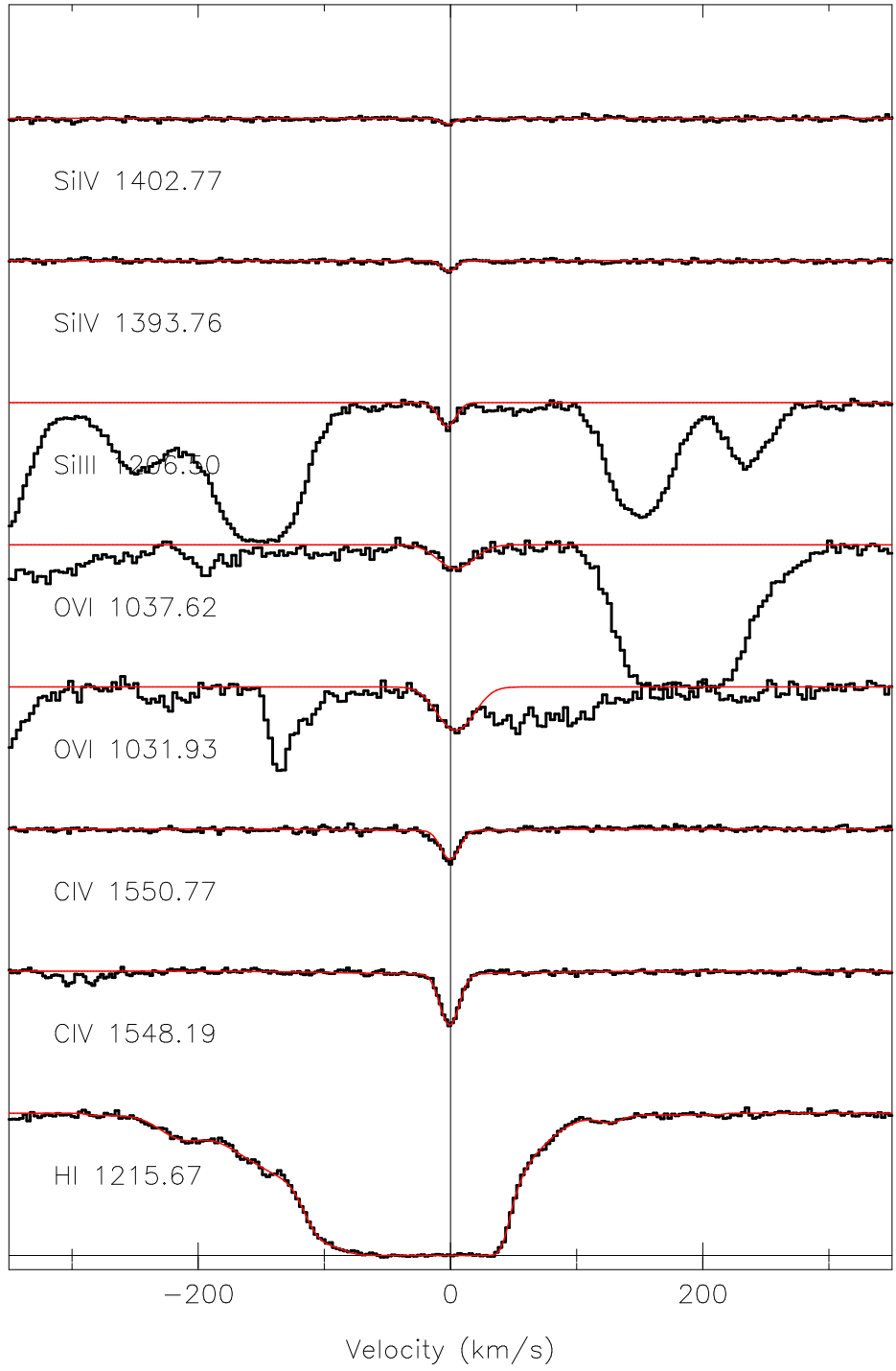}
\caption{ Stacked velocity plot of the $z=2.379$ in Q1700+6416 
(Described in Section \ref{sec_1700_2.37} of the text).}
\end{figure}

\begin{figure}
\figurenum{15}
\epsscale{0.75}
\plotone{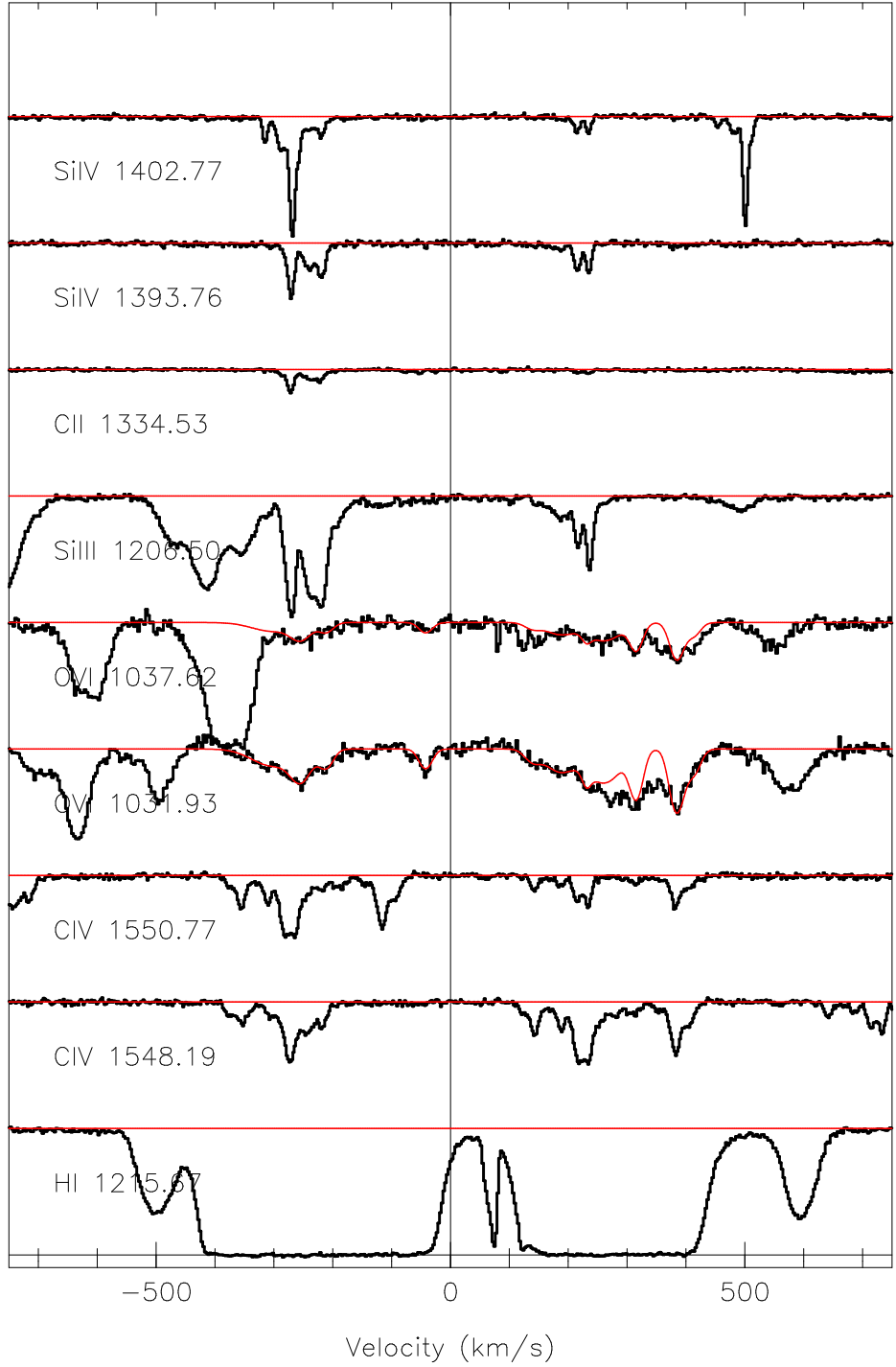}
\caption{ Stacked velocity plot of the $z=2.433/2.439$ system in 
Q1700+6416 (Described in Section \ref{sec_1700_2.43} of the text).
The \ovi 1032\ang profile has been adjusted to remove an interloping
\lyb line at $z=2.463$ (+300 km/s), to better illustrate the
correspondence of the \ovi doublet profiles.}
\end{figure}

\begin{figure}
\figurenum{16}
\epsscale{0.75}
\plotone{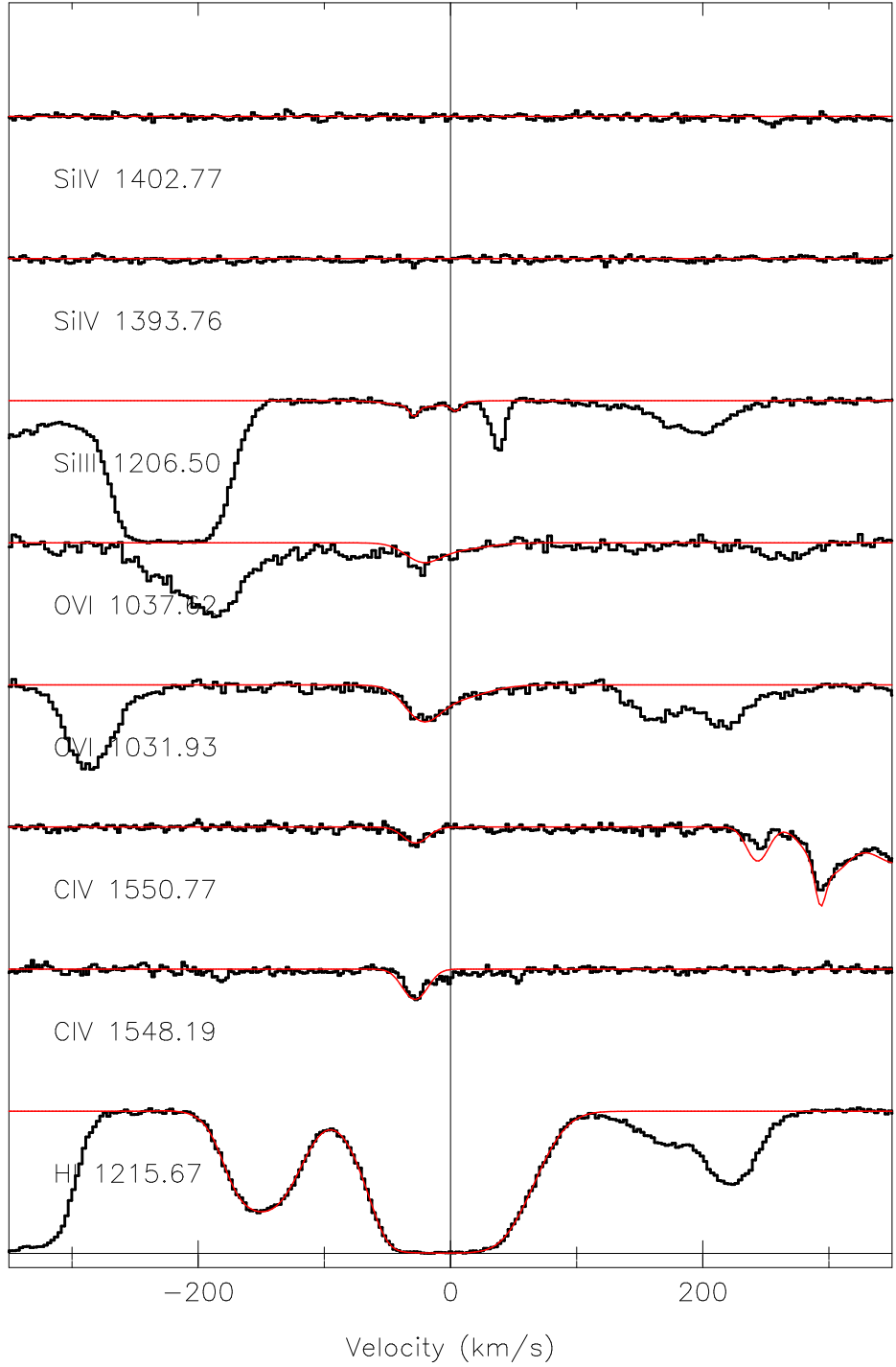}
\caption{ Stacked velocity plot of the $z=2.568$ system in 
Q1700+6416 (Described in Section \ref{sec_1700_2.56} of the text).}
\end{figure}

\begin{figure}
\figurenum{17}
\epsscale{0.75}
\plotone{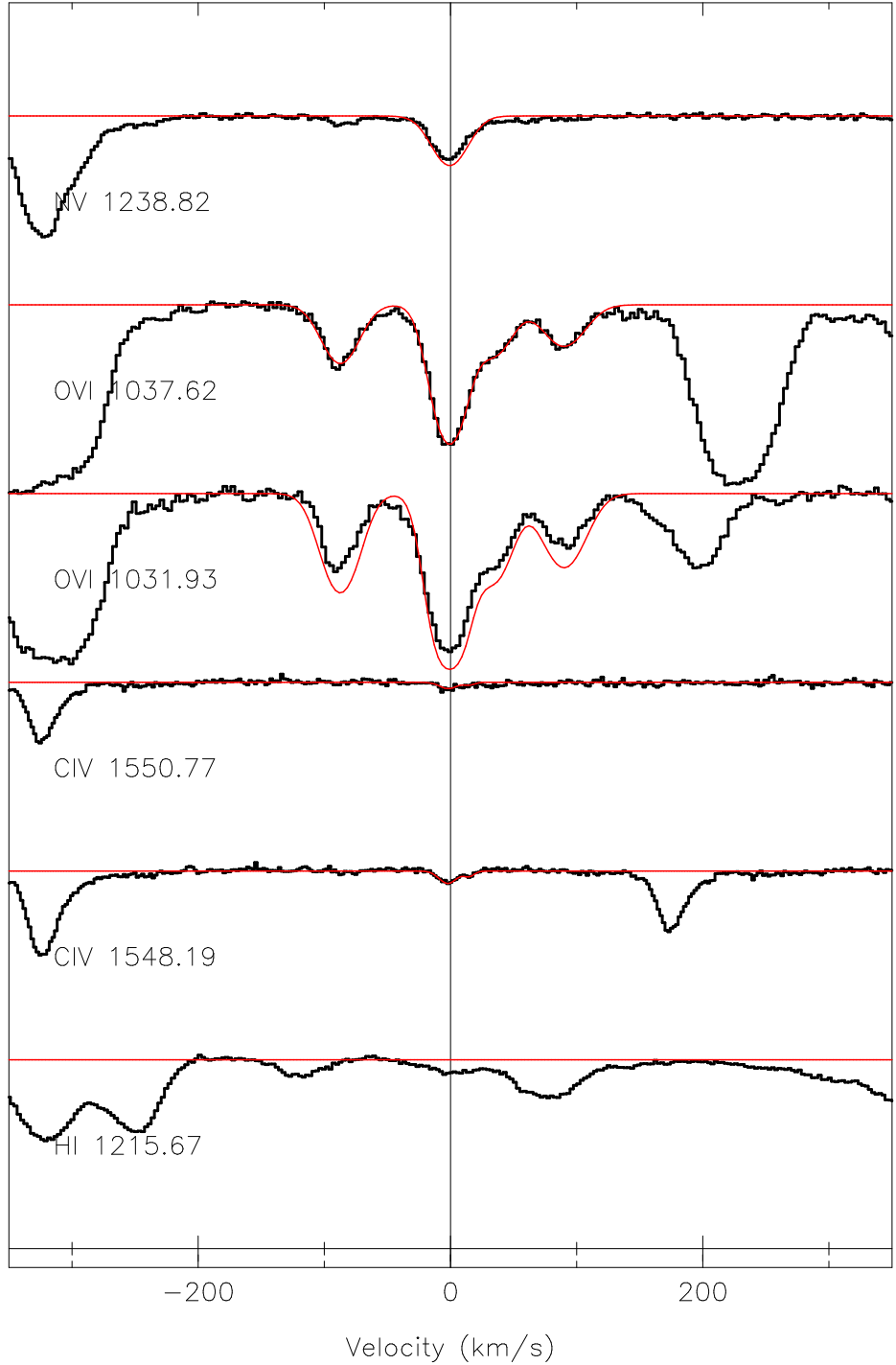}
\caption{ Stacked velocity plot of the $z=2.716$ system in 
Q1700+6416 (Described in Section \ref{sec_1700_2.71} of the text).
The model fit shown in the figure illustrates the expected strength of
the \ovi 1032\ang line based on a fit to the 1037\ang profile.
Because the observed profile does not match the expected doublet
ratio, we have surmised that this absorber is ejected from the
background QSO and only partially covers the central engine.  It has
been excluded from the cosmological statistics.}
\end{figure}

\begin{figure}
\figurenum{18}
\epsscale{0.75}
\plotone{f18.eps}
\caption{ Stacked velocity plot of the $z=2.745$ system in 
Q1700+6416 (Described in Section \ref{sec_1700_2.74} of the text).
This system has been excluded from the cosmological statistics sue to
its proximity to the background QSO.}
\end{figure}

\begin{figure}
\figurenum{19}
\label{fig_pathlength}
\plotone{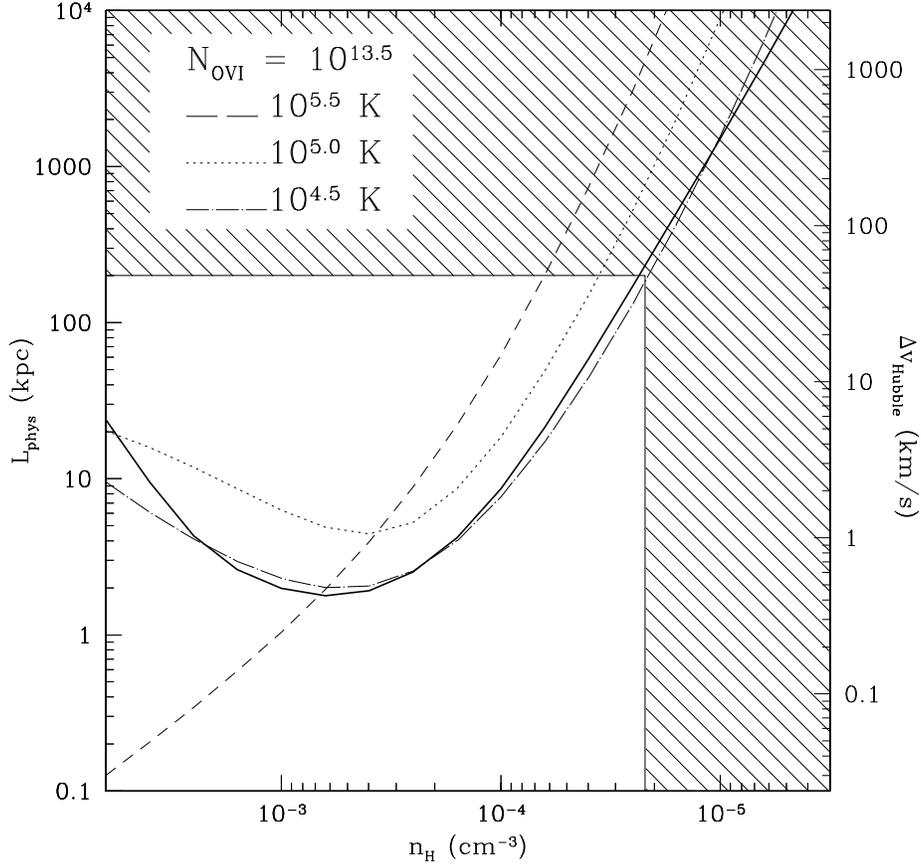}
\caption{\ovi absorption pathlength for a line with
$N_{\movi}=10^{13.5}$ \pcmsq, for different gas number densities,
derived from CLOUDY calculations.  The right axis shows the expected 
broadening $\Delta v=H_{(z=2.5)}L$ for lines corresponding to
structures of different sizes.  The solid line shows the
photoionization equilibrium value; other lines indicate solutions at
fixed temperatures as indicated.  The shaded portions of the diagram
are ruled out by the observed linewidths, which have an upper limit of
$48$ km/s.  The excluded region in density is derived from the most
conservative model curve (the photoionization equilibrium model).
Note that the density scale is shown decreasing to the right, for
consistency with Figures \ref{fig_civ_siiv} and \ref{fig_ovi_civ}.}
\end{figure}

\begin{figure}
\figurenum{20}
\label{fig_b_temp}
\plotone{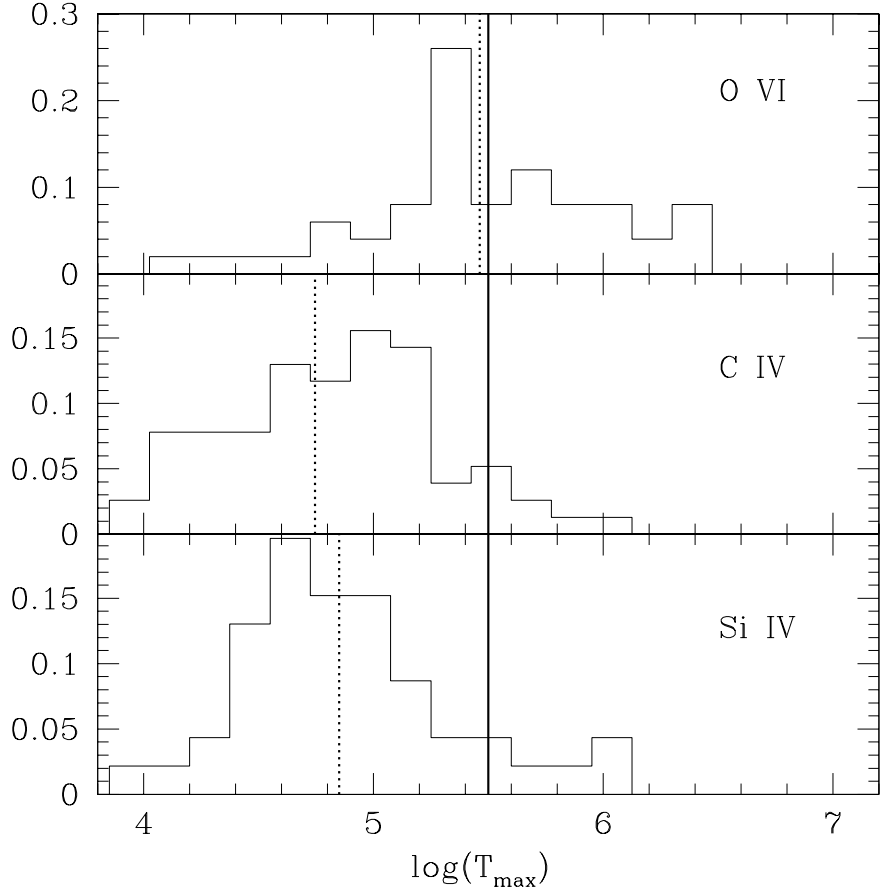}
\caption{Histogram of upper limits on the temperature for all \ovi,
\civ, and \siiv components observed in our systems, assuming
completely thermal line broadening.  The solid vertical line drawn at
$\log(T_{\rm max})=5.5$ indicates the approximate temperature where
collisional production of \ovi should peak.  The three vertical dashed
lines indicate the mean value of $T_{\rm max}$ for each ion. The \civ 
and \siiv temperatures are clearly too low to support the collisional
ionization of \ovi, and appear to be contained in a diferent phase than the
\ovi.  The distribution of $T_{\rm max}$ for \ovi tentatively appears 
to favor hotter temperatures characteristic of collisionally ionized
\ovi, though the true distribution may be lower due to non-thermal
(e.g. turbulent) line broadening.
}
\end{figure}

\begin{figure}
\figurenum{21}
\label{fig_civ_siiv}
\plotone{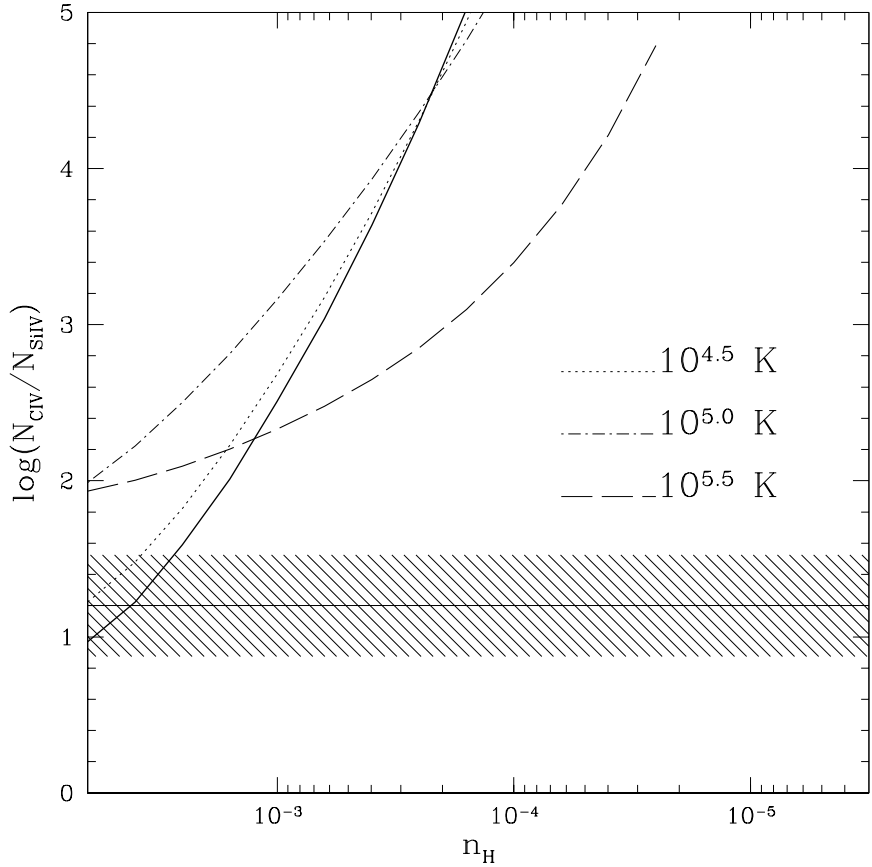}
\caption{ \civ/\siiv column density ratio predicted by CLOUDY as a
function of gas number density (Note that the density scale {\it decreases}
to the right, corresponding to an increasing ionization parameter).  
The solid line is the photoionization
equilibrium solution, other lines are at fixed temperatures.
The shaded region indicates
the $\pm 1\sigma$ range in the ratio measured in \ovi systems where 
both \siiv and \civ were detected.  Models with $T>10^{4.6}$ K are
ruled out at any density; the most likely model is a photoionized gas
with $n_H\ge 4\times 10^{-3}$ at $T=1-5\times 10^4$ K.  This
temperature is consistent with measurements of the \civ and \siiv
linewidths, when one uses both ions to solve for the thermal and 
non-thermal contributions to line broadening simultaneously.}
\end{figure}

\begin{figure}
\figurenum{22}
\label{fig_ovi_civ}
\plotone{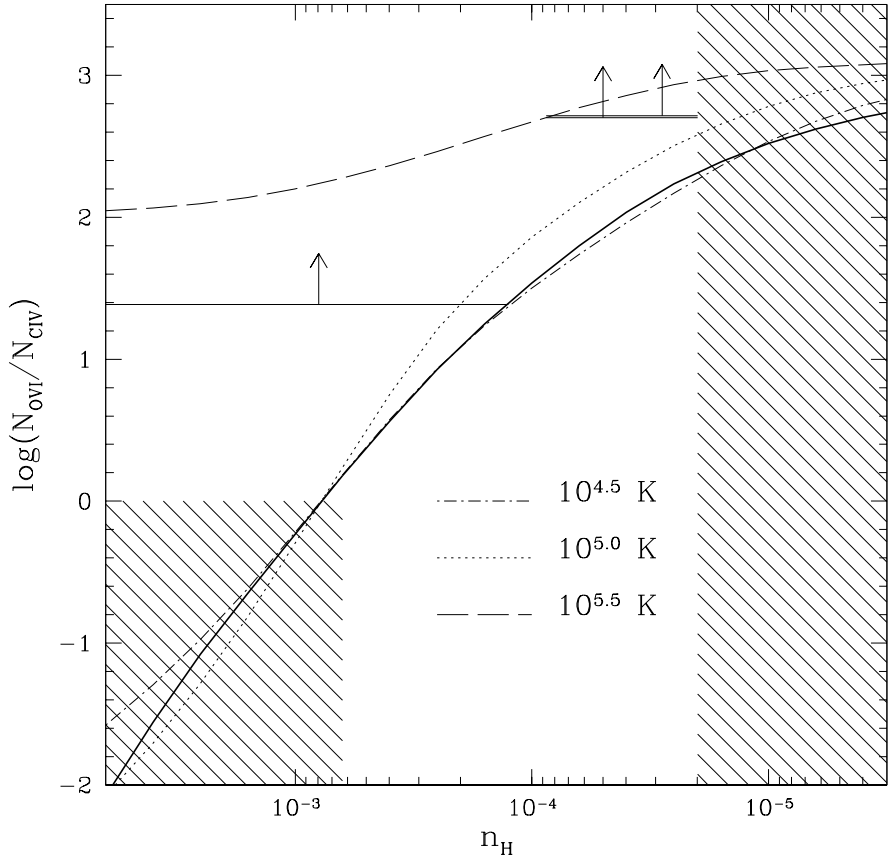}
\caption{ \ovi/\civ column density ratio predicted by CLOUDY as a
function of gas number density (Note that the density scale decreases
to the right, as in Figure \ref{fig_civ_siiv}).  
Solid line represents a photoionization
equilibrium solution, other lines are at fixed temperatures
as indicated.
Shaded region at right is excluded based
on the pathlength arguments presented in Section
\ref{sec_pathlength}.  Shaded region at lower left is excluded based
on the observed properties of \civ and \siiv.  Horizontal solid lines
indicate upper limits on the ratio from the two systems where a measurement
could reliably be made.  For most systems, an accurate measurement
could not be made because of confusion in the \civ profile with
absorption from the cooler Carbon-Silicon phase.}
\end{figure}

\begin{figure}
\figurenum{23}
\label{fig_tpcf}
\plotone{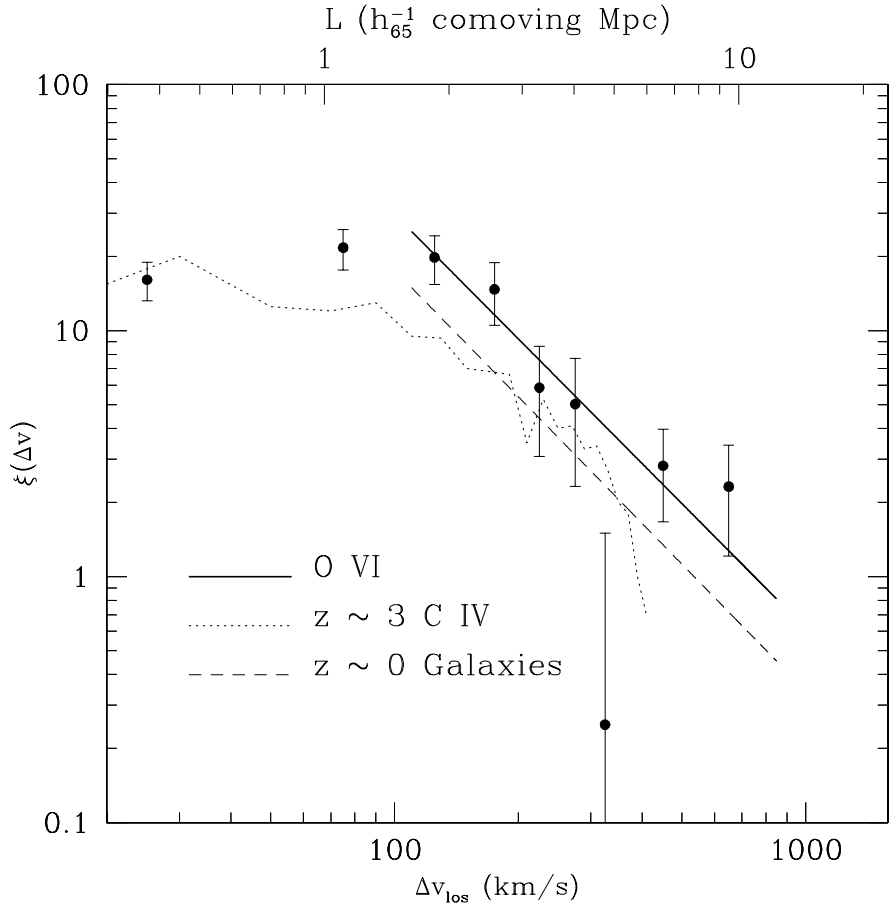}
\caption{ Two-point correlation funtion of \ovi clouds along the line
of sight for all intergalactic systems in the survey.  
The errorbars represent $1\sigma$ uncertainties due to finite sample size.
The solid line denotes our best fit power law, given in Equation
\ref{eqn_powerlaw}.  For comparison, the \civ correlation function 
from \citet{rauch1996} is shown as a dotted line, and the correlation
function of local galaxies \citep{loveday1995} is shown as a dashed line.}
\end{figure}

\end{document}